\documentclass{aa}
\usepackage{graphicx}
\begin{document}

\title{Magnetic fields of HgMn stars\thanks
{Based on observations obtained at the European Southern Observatory (ESO programmes 076.D-0169(A),
076.D-0172(A), 084.D-0338(A),
085.D-0296(A), 085.D-0296(B), 087.D-0049(A), 088.D-0284(A)), SOFIN observations at the 
2.56 m Nordic Optical Telescope on La Palma, and observations obtained with the CORALIE Echelle Spectrograph on the 1.2 m Euler Swiss telescope
on La Silla, Chile.}}

\author{
S. Hubrig\inst{1}
\and
J.~F.~Gonz\'alez\inst{2}
\and
I.~Ilyin\inst{1}
\and
H.~Korhonen\inst{3,4}
\and
M.~Sch\"oller\inst{5}
\and
I.~Savanov\inst{6}
\and
R.~Arlt\inst{1}
\and
F.~Castelli\inst{7}
\and
G.~Lo~Curto\inst{5}
\and
M.~Briquet\inst{8,9}\thanks{F.R.S.-FNRS Postdoctoral Researcher, Belgium}
\and
T.~H.~Dall\inst{5}
}

\institute{
Leibniz-Institut f\"ur Astrophysik Potsdam (AIP), An der Sternwarte 16, 14482 Potsdam, Germany\\
\email{shubrig@aip.de}
\and
Instituto de Ciencias Astronomicas, de la Tierra, y del Espacio (ICATE), 5400 San Juan, Argentina
\and
Niels Bohr Institute, University of Copenhagen, Juliane Maries Vej 30, 2100, K\o{}benhavn, Denmark
\and
Centre for Star and Planet Formation, Natural History Museum of Denmark, University of Copenhagen, 
Øster Voldgade 5-7, DK-1350 Denmark 
\and
European Southern Observatory, Karl-Schwarzschild-Str.\ 2, 85748 Garching bei M\"unchen, Germany
\and
Institute of Astronomy, Russian Academy of Sciences, Pyatnitskaya 48, Moscow 119017, Russia
\and
Instituto Nazionale di Astrofisica, Osservatorio Astronomico di Trieste, via Tiepolo 11, 34143 Trieste, 
Italy 
\and
Instituut voor Sterrenkunde, K.U.~Leuven, Celestijnenlaan 200D, 3001, Leuven, Belgium 
\and
Institut d'Astrophysique et de G\'eophysique Universit\'e de Li\`ege, All\'ee du 6 Ao\^ut 17, B-4000 Li\`ege, Belgium
}

\abstract
 % context heading (optional)
{
The frequent presence of weak magnetic fields on the surface of spotted late-B stars with HgMn peculiarity
in binary systems has been controversial during the two last decades.
Recent studies of magnetic fields in these stars using the least-squares deconvolution (LSD) technique 
have failed to detect magnetic fields, indicating an upper limit on the longitudinal field  
between 8 and 15\,G. 
%although the authors were aware of the 
%inhomogeneous distribution  of elements on the surface of HgMn stars, 
In these LSD studies, assumptions 
were made that all spectral lines are 
identical in shape and can be described by a scaled mean profile. 
}  
{
We re-analyse the 
available spectropolarimetric material by applying the moment technique on spectral lines of 
inhomogeneously 
distributed elements separately. Furthermore, we present new
determinations of the mean longitudinal magnetic field for the HgMn star HD\,65949 and the hotter analog of 
HgMn stars, the PGa star HD\,19400, using FORS\,2 installed at the VLT. We also give
new measurements of the eclipsing system AR\,Aur with a primary star of HgMn peculiarity which were obtained 
with the SOFIN spectropolarimeter installed at the Nordic Optical Telescope. 
}
{ 
We downloaded from the European Southern Observatory (ESO) archive the publically available  
HARPS spectra for eight HgMn stars and one normal and one superficially normal B-type star obtained
in 2010. Out of this sample, three HgMn stars belong 
to spectroscopic double-lined systems. 
The application of the moment technique to the HARPS and SOFIN spectra allowed us to study 
the presence of the longitudinal magnetic field, the crossover effect, and quadratic magnetic fields. 
Results for the HgMn star HD\,65949 and the PGa star HD\,19400 are based on a linear regression 
analysis of low-resolution spectra obtained with FORS\,2 in spectropolarimetric mode.
}
{
Our measurements of the magnetic field with the moment technique using spectral lines of several 
elements separately reveal the presence of a weak longitudinal magnetic field, a quadratic magnetic field,
and the crossover effect
%of the order of tens  of gauss 
on the surface of several HgMn stars as well as normal and superficially normal B-type stars. 
Furthermore, our analysis suggests the 
existence of intriguing correlations between the strength of the magnetic field, abundance anomalies, and binary 
properties. The results are discussed in the context of possible mechanisms responsible for the 
development of the element patches and complex magnetic fields on the surface of late B-type stars.
% of the order of a few kG.
}
{}


\titlerunning{Magnetic fields of HgMn stars}
\authorrunning{S.\ Hubrig et al.}
\maketitle

%________________________________________________________________

\section{Introduction}
\label{sect:intro}

The origin of abundance anomalies observed in late B-type stars with HgMn peculiarity is still poorly 
understood. 
Over the last few years, we have performed 
extensive spectroscopic studies of both single late B-type stars and of spectroscopic binaries (SB) with 
late B-type primaries  (spectral types B7--B9)
with the goal of understanding why the vast majority of these stars exhibit in their 
atmospheres certain chemical abundance anomalies, i.e.\ large excesses of P, Mn, Ga, Br, Sr, Y, Zr, 
Rh, Pd, Xe, Pr, Yb, W, Re, Os, Pt, Au, and Hg, and 
underabundances of He, Al, Zn, Ni, and Co (e.g. Castelli \& Hubrig \cite{Castelli2004a}). Strong isotopic 
anomalies were detected for the chemical elements Ca, Pt, and Hg, with patterns changing from one star to 
the next (Hubrig et al.\ \cite{Hubrig1999a}; Castelli \& Hubrig \cite{Castelli2004b}; 
Cowley et al.\ \cite{Cowley2010}).
Observationally, these stars are characterised by low rotational 
velocities ($\langle v\,\sin i\rangle \leq 29$~km\,s$^{-1}$, 
Abt et al.\ \cite{Abt1972}). The number of these chemically peculiar stars, 
usually called HgMn stars, decreases with increasing rotational
velocity (Wolff \&\ Wolff \cite{Wolff1974}). 
%Evidence that stellar rotation does affect abundance anomalies in HgMn stars is provided by the rather 
%sharp cutoff in such anomalies at a projected rotational velocity 
%of 70--80~km\,s$^{-1}$ (Hubrig \& Mathys 1996).

More than two thirds of the HgMn stars are known to belong to spectroscopic 
binaries (Hubrig \& Mathys \cite{Hubrig1995}), with a preference of orbital 
periods ranging from 3 to 20 days.
It is striking that the inspection of SB systems with a late B-type 
primary in the 9$^{\rm th}$ Catalogue of Spectroscopic Binary 
Orbits (Pourbaix et al.\ \cite{Pourbaix2009}) indicates a strong correlation 
between the HgMn peculiarity and membership in a binary system
among bright, well-studied SB systems with late B-type, slowly rotating 
\mbox{($v\,\sin i<\,70~{\rm km\,s}^{-1}$)} primaries with an apparent 
magnitude of up to $V\approx7$  and orbital periods between 3 and 
20\,d. With the exception of HR\,7241, all 21~systems have a primary with 
a HgMn peculiarity. Based on this fact, it is very likely that the 
majority of slowly rotating late B-type stars formed in binary 
systems with certain orbital parameters become HgMn stars. Consequently, careful studies of these peculiar 
stars are important for the general understanding of B-type star 
formation in binary systems. According to Abt \& Snowden (\cite{Abt1973}), close-binary formation
is inhibited when strong magnetic fields are present.
They also suggest that close binaries dissipate the magnetic field.

The aspect of inhomogeneous distribution of some 
chemical elements over the surface of HgMn stars was first discussed by 
Hubrig \& Mathys (\cite{Hubrig1995}). 
From a survey of HgMn stars in close SBs, it was suggested that some chemical elements might be 
inhomogeneously distributed on the surface, with, in particular, preferential concentration of Hg along 
the equator. 
The first definitively identified spectrum variability 
was reported for the binary HgMn star $\alpha$~And by
%Wahlgren, Ilyin \& Kochukhov (\cite{Wahlgren2001})
Wahlgren et al.\ (\cite{Wahlgren2001})
and Adelman et al.\ (\cite{Adelman2002}), who showed that 
the spectral variations of the 
\ion{Hg}{ii} line at $\lambda$3984 discovered in high-dispersion spectra are not due to the orbital motion of the 
companion, but produced by the combination of the 2.8\,d period of rotation of the primary and a non-uniform 
surface distribution of mercury, which is concentrated in the equatorial region. Their results are 
in good correspondence 
with those of Hubrig \& Mathys (\cite{Hubrig1995}). The variability of the \ion{Hg}{ii} line 
at $\lambda$3984 was 
interpreted with a Doppler imaging code, revealing high-contrast mercury spots located along the 
rotational equator. Using a Doppler imaging reconstruction of spectroscopic time series obtained 
over seven consecutive years, Kochukhov et al.\ (\cite{Kochukhov2007}) suggested the presence of a secular 
evolution of the mercury distribution. On the other hand, the work of other authors has proved that not only 
mercury abundance appears distributed in patches over the stellar surface. Almost all other elements,
most typical Ti, Cr, Fe, Mn, Sr, Y, and Pt, are concentrated in spots of diverse size, and different elements
exhibit  different abundance distributions across the stellar surface (e.g. Hubrig et al.\ \cite{Hubrig2006a}; 
Briquet et al.\ \cite{Briquet2010}; Makaganiuk et al.\ \cite{Makaganiuk2011a}). 
Moreover, an evolution of the element abundance spots at different time scales was discovered
in two additional HgMn stars.  Briquet et al.\ (\cite{Briquet2010}) reported the presence of 
dynamical spot evolution over a couple of 
weeks for the SB1 system HD\,11753, while Hubrig et al.\ (\cite{Hubrig2010}) detected
a secular element evolution in the double-lined eclipsing binary AR\,Aur.
Importantly, recent results (e.g. Nu{\~n}ez et al.\ \cite{Nunez2011}; Hubrig et al.\ \cite{Hubrig2011}) 
show that line-profile 
variability of various elements caused by a non-uniform abundance distribution is a general characteristic 
of HgMn stars, rather than an exception. 
%This variability is caused by an inhomogeneous chemical element distribution, and 
%implies that most HgMn stars present a non-uniform distribution of one or more chemical elements. 

Typically, inhomogeneous chemical abundance distributions are 
observed only on the surface of magnetic chemically peculiar 
stars with large-scale organised magnetic fields (Ap and Bp stars). In these 
stars, the abundance distribution of certain elements is 
non-uniform and non-symmetric with respect to the rotation axis. 
Numerous studies of Ap and Bp stars have revealed a kind of symmetry
between the topology of the magnetic field and the element distribution.
Thus, the structure of the magnetic field can be studied by measurements of 
the magnetic field, using spectral lines of each element separately.
%expected, the method of using all element spectral lines together is not advisable and
% leads to doubtful results.  
Although strong large-scale magnetic fields have not generally been
found in HgMn stars, it has never been ruled out that these 
stars might have tangled magnetic fields of the order of a few 
thousand Gauss with only very weak net longitudinal components.
Bychkov et al.\ (\cite{Bychkov2009}) compiled all published longitudinal magnetic field measurements
of different groups of chemically peculiar stars and concluded that the group of HgMn stars possesses the second 
weakest fields after the group of Am stars.

In the last few years, a number of  attempts to detect mean longitudinal
magnetic fields in HgMn stars have been made by several authors using the line 
addition technique, the least-squares deconvolution (LSD), most recently by Makaganiuk et al.\ 
%(e.g., Makaganiuk et al.\ \cite{Makaganiuk2011b}, \cite{mak2011c}). 
(e.g. \cite{Makaganiuk2011a}, \cite{Makaganiuk2011b}, \cite{mak2011c}). 
A high level of precision, from a few to tens of Gauss, is achieved through 
application of the LSD technique
%of magnetic field determination
(Donati et al.\ \cite{Donati1997}), which combines 
hundreds of spectral lines of various elements. This technique assumes that all spectral lines 
are identical in shape and can be described by a 
scaled mean profile. However, the lines of 
different elements with different abundance distributions across the stellar 
surface sample the magnetic field in different manners. Combining them as is 
done with the LSD technique may lead to the dilution of the magnetic 
signal or even to its (partial) cancellation, if enhancements of different 
elements occur in regions of opposite magnetic polarities. 
%The average line profiles 
%are calculated for thousands of spectral lines of all elements all together.
%%without consideration of the inhomogeneous element distribution on the surface of HgMn stars. 
Using this technique, Makaganiuk et al.\ (\cite{Makaganiuk2011b})
analysed HARPS spectra for a sample of HgMn and normal B-type stars and reported no 
detection at a 3$\sigma$ level in any of the studied targets.

Strangely enough, although the authors were aware of the 
inhomogeneous distribution of the elements on the surface of HgMn stars, no 
analysis has been done on the lines of individual elements separately. 
%Since a kind of symmetry between the topology of magnetic fields and the element distribution is
%expected, the method of using all element spectral lines together is not advisable and
% leads to doubtful results.  
Only in the most recent Makaganiuk et al.\ (\cite{mak2012}) study did the authors decide to constrain the 
longitudinal magnetic field of HD\,11753 by computing mean profiles for Y, Ti, and Cr, which 
show different spot distributions on the stellar surface. Their results 
indicate that the upper limit for the strength of the magnetic fields can reach about 20--30\,G, with a 
typical error bar from 8\,G to 15\,G.

\begin{table}
\centering
\caption{
List of the studied targets.
}
\label{tab:targetlist}
\centering
\begin{tabular}{lcrcc}
\hline
\hline
\multicolumn{1}{c}{Object \rule{0pt}{2.6ex}} &
\multicolumn{1}{c}{Other} &
\multicolumn{1}{c}{V} &
\multicolumn{1}{c}{Spectral} &
\multicolumn{1}{c}{Instr.} \\
\multicolumn{1}{c}{Name} &
\multicolumn{1}{c}{Identifier} &
 &
\multicolumn{1}{c}{Type} &
\\
\hline
HD\,11753     & $\phi$\,Phe & 5.11 & B9V   & HARPS \\
HD\,19400     & $\theta$\,Hyi & 5.50 & B3V   & FORS\,2 \\
HD\,27376     & 41\,Eri     & 3.55 & B9V, SB2   & HARPS \\
HD\,32964    & 66\,Eri       & 5.10  &  B9V, SB2    & HARPS \\
HD\,33904   & $\mu$\,Lep    & 3.28  & B9IV     & HARPS \\
HD\,34364   & AR\,Aur        & 6.14  & B9V, SB2    & SOFIN \\
HD\,53244   & $\gamma$\,CMa   & 4.10  & B8II    & HARPS \\
HD\,65949   & CPD$-$60\,966    & 8.37  & B8/B9    & FORS\,2 \\
HD\,78316   & $\kappa$\,Cnc   & 5.24  & B8III, SB2    & HARPS \\
HD\,101189   &HR\,4487        & 5.14  & B9IV    & HARPS \\
HD\,221507   & $\beta$\,Scl   & 4.37  & B9.5IV    & HARPS \\
\hline
\multicolumn{5}{c}{(Superficially) normal B-type stars}\\
\hline
HD\,179761  &V1288\,Aql & 5.14 & B8II-III     & HARPS \\
HD\,209459  & 21\,Peg    & 5.82 & B9.5V      & HARPS \\
\hline
\end{tabular}
%\begin{flushleft}
%Notes:
\tablefoot{
Spectral types and visual magnitudes are taken from SIMBAD.
}
%\end{flushleft}
\end{table}

To test if the use of the LSD technique may indeed account for the non-detection of magnetic fields
in HgMn stars, we decided to re-analyse the HARPS spectropolarimetric material that recently
became publically available in the ESO archive.
%Since a part of high-quality HARPS spectropolarimetric material used in studies of Makaganiuk et al.\ became 
%publicly available in the last months, we re-analysed a number of spectra 
In this work, we use a completely
different approach for the measurements of magnetic fields, 
namely, the moment technique developed by Mathys (e.g. \cite{Mathys1991}, \cite{Mathys1995a}, \cite{Mathys1995b}).
This technique allows us not only to determine the mean longitudinal magnetic field, but 
also to prove the presence of the crossover effect and quadratic magnetic fields.
This information cannot be obtained from the LSD technique, as it assumes  that all spectral lines are 
identical in shape and can be described by a scaled mean profile. 
Furthermore, we present five new spectropolarimetric observations of the eclipsing 
system AR\,Aur with a primary star of HgMn peculiarity, obtained 
with the echelle spectrograph SOFIN at the Nordic Optical Telescope at the end of 2010.
The variability and spot distribution of this star has been 
intensively studied by our group in the few last years (Hubrig et al.\ \cite{Hubrig2006a}, \cite{Hubrig2011}). 
In addition, a new set of polarimetric spectra
was obtained for  the HgMn star HD\,65949 and the hotter analog of 
HgMn stars, the PGa star HD\,19400, using FORS\,2 installed at the VLT.
All targets discussed in this work are presented in Table~\ref{tab:targetlist}
together with their visual magnitudes and spectral types.

In Sect.~\ref{sect:obs}, we describe the observations, data reduction, and methods of our magnetic field
measurements.  
The stellar characteristics and the  magnetic field measurements for the individual targets are reviewed 
in Sect.~\ref{sect:indivi}. 
Finally, in Sect.~\ref{sect:disc}, we discuss our results in the context of possible mechanisms at play, which can be 
considered responsible for the development of the inhomogeneous abundance distribution and the presence of 
weak magnetic fields.
% of present theoretical MHD calculations.

\section{Observations and magnetic field measurements}
\label{sect:obs}

\subsection{HARPS observations}

\begin{table}
\centering
\caption{
Logbook of the spectropolarimetric observations.
%Dates and S/N values for spectropolarimetric observations.
}
\label{tab:s/n}
\centering
\begin{tabular}{llr}
\hline
\hline
\multicolumn{1}{c}{Object} &
%\rule{0pt}{2.6ex}} &
\multicolumn{1}{c}{MJD} &
\multicolumn{1}{c}{S/N}  \\
%\multicolumn{1}{c}{Spectral} \\
%\multicolumn{1}{c}{Name} &
%\multicolumn{1}{c}{Identifier} &
% &
%\multicolumn{1}{c}{Type} \\
\hline
\multicolumn{3}{c}{HARPS observations}\\
\hline
HD\,11753    &55199.0975  &332 \\
             &55200.0549  &299\\
             &55201.0296  &395\\
             &55202.0438  &433\\
             &55203.0356  &348\\
             &55204.0228  &430\\
             &55205.0518  &353\\
             &55206.0271  &406\\
             &55207.0207  &476\\
             &55209.0147  &379\\
             &55210.0368  &564\\
41\,Eri      &55201.2687  &507\\
             &55210.2561  &422\\
             &55212.2414  &589\\
             &55213.2305  &411\\
66\,Eri      &55202.2262  &245\\
             &55203.2457  &299\\
             &55204.2384  &320\\
             &55205.2570  &185\\
             &55206.2375  &304\\
             &55207.2363  &315\\
             &55209.2426  &261\\
             &55210.2476  &216\\
             &55211.2523  &241\\
             &55212.2290  &380\\
HD\,33904    &55204.2676  &600 \\ 
HD\,53244    &55204.2799  &477 \\ 
$\kappa$\,Cnc  &55202.3038  &230 \\ 
             &55211.2873  &339\\
HD\,101189   &55201.3629  &179\\
%&54982.4744  &51 \\ 
             &55201.3629  &179\\
HD\,179761   &55319.3267  &123 \\ 
HD\,209459   &55417.0999  & 80 \\ 
             &55417.1395  & 87\\
             &55421.2527  &257\\
HD\,221507   &55210.0485  &349  \\ 
\hline
\multicolumn{3}{c}{SOFIN observations}\\
\hline
AR\,Aur      &55544.1347  & 288 \\
%&55195.1390  & 332\\ 
             &55553 2003  & 310\\
             &55554.0300  & 365 \\
             &55555.1200  & 342\\
             &55556.0350  & 346 \\
\hline
\multicolumn{3}{c}{FORS\,2 observations}\\
\hline
HD\,19400  & 55845.2952& 1604 \\
           & 55935.1094& 1322 \\
HD\,64949  & 54108.1771&  794 \\
           & 54433.3662&  876 \\
           & 55686.0703& 1304 \\
           & 55688.0404&  911 \\
%%$-$182(34)(116/32) \\
\hline
\end{tabular}
%%\begin{flushleft}
%%Notes:
%\tablefoot{
%Spectral types and visual magnitudes are taken from SIMBAD.
%}
%%\end{flushleft}
\end{table}

The majority of the spectra analyzed in this work were obtained with the HARPSpol polarimeter 
(Snik et al.\ \cite{Snik2011}) feeding the HARPS spectrometer 
(Mayor et al.\ \cite{Mayor2003}) at the ESO 3.6\,m telescope on La Silla. 
We downloaded from the ESO archive the publically available multi-epoch and single-epoch 
spectra for eight HgMn stars, one normal B-type star, and one superficially normal B-type star obtained
with HARPSpol in 2010. 
Out of this sample, three HgMn stars belong 
to spectroscopic double-lined systems. The study of magnetic fields in one normal star and one superficially 
normal B-type 
star is of special interest because they were reported as weakly magnetic in previous polarimetric studies. 
%Five observations on different nights were obtained in February 2011. 
 All spectra have a resolving power of $R = 115\,000$ and a signal-to-noise ratio (S/N) between 80 for 
HD\,209459 and 589 for 41\,Eri.  
%allowing to obtain measurement uncertainties as small as a few Gauss. 
 The information on the individual observations, including the MJD dates and the S/N values, is 
given in Table~\ref{tab:s/n}. To distinguish between SB1 and SB2 
systems, we will call the SB1 systems with their HD number and
use the other identifiers given in the second column of Table~\ref{tab:targetlist} for the SB2 systems.

\begin{table*}
\caption[]{
Measurements of the mean longitudinal magnetic field using sets of lines belonging to different elements.
All quoted errors are 1$\sigma$ uncertainties.
}
\label{tab:log_meas}
\begin{center}
\begin{tabular}{lrr@{$\pm$}lr@{$\pm$}lr@{$\pm$}lr@{$\pm$}lr@{$\pm$}lr@{$\pm$}lr@{$\pm$}lr@{$\pm$}l}
\hline \hline\\[-7pt]
\multicolumn{1}{c}{MJD} &
\multicolumn{1}{c}{Phase} &
\multicolumn{2}{c}{$\left<B_{\rm z}\right>_{\rm Ti}$} &
\multicolumn{2}{c}{$\left<B_{\rm z}\right>_{\rm Ti,n}$} &
\multicolumn{2}{c}{$\left<B_{\rm z}\right>_{\rm Cr}$} &
\multicolumn{2}{c}{$\left<B_{\rm z}\right>_{\rm Cr,n}$} &
\multicolumn{2}{c}{$\left<B_{\rm z}\right>_{\rm Fe}$} &
\multicolumn{2}{c}{$\left<B_{\rm z}\right>_{\rm Fe,n}$} &
\multicolumn{2}{c}{$\left<B_{\rm z}\right>_{\rm Y}$} &
\multicolumn{2}{c}{$\left<B_{\rm z}\right>_{\rm Y,n}$} \\
&
&
\multicolumn{2}{c}{[G]} &
\multicolumn{2}{c}{[G]} &
\multicolumn{2}{c}{[G]} &
\multicolumn{2}{c}{[G]} &
\multicolumn{2}{c}{[G]} &
\multicolumn{2}{c}{[G]} &
\multicolumn{2}{c}{[G]} &
\multicolumn{2}{c}{[G]} \\
\hline\\[-7pt]
\multicolumn{18}{c}{HD\,11753} \\
\hline\\[-7pt]
  55203.0356 & 0.098 &     0 & 16 & \multicolumn{2}{c}{} &    12 & 16 & \multicolumn{2}{c}{} &    18 & 11 & \multicolumn{2}{c}{} &  $-$1 & 20 & \multicolumn{2}{c}{} \\
  55204.0228 & 0.203 & $-$39 & 12 &  0 & 13              & $-$14 & 11 & \multicolumn{2}{c}{} & $-$24 & 13 & \multicolumn{2}{c}{} & $-$48 & 11 & $-$17 & 13 \\
  55205.0518 & 0.308 &  $-$3 & 10 & \multicolumn{2}{c}{} & $-$18 & 13 & \multicolumn{2}{c}{} &  $-$7 & 11 & \multicolumn{2}{c}{} &     9 & 15 & \multicolumn{2}{c}{} \\
  55206.0271 & 0.413 &     8 &  9 & \multicolumn{2}{c}{} & $-$11 & 16 & \multicolumn{2}{c}{} &  $-$2 & 11 & \multicolumn{2}{c}{} &    10 & 15 & \multicolumn{2}{c}{} \\
  55207.0207 & 0.518 &     1 & 10 & \multicolumn{2}{c}{} &     8 & 11 & \multicolumn{2}{c}{} &    12 & 10 & \multicolumn{2}{c}{} &  $-$3 & 10 & \multicolumn{2}{c}{} \\
  55199.0975 & 0.678 &     7 & 12 & \multicolumn{2}{c}{} &    20 & 15 & \multicolumn{2}{c}{} &    18 & 16 & \multicolumn{2}{c}{} &     4 & 15 & \multicolumn{2}{c}{} \\
  55209.0147 & 0.729 &    12 & 11 & \multicolumn{2}{c}{} &    20 &  8 & \multicolumn{2}{c}{} &     8 & 11 & \multicolumn{2}{c}{} &    18 & 19 & \multicolumn{2}{c}{} \\
  55200.0549 & 0.783 &    26 &  9 & \multicolumn{2}{c}{} &    63 & 19 & $-$12 & 17           &    25 &  7 & 10 & 8              &    45 & 19 & \multicolumn{2}{c}{} \\
  55210.0368 & 0.832 &     9 & 10 & \multicolumn{2}{c}{} &  $-$9 & 16 & \multicolumn{2}{c}{} &    15 & 11 & \multicolumn{2}{c}{} &     4 & 11 & \multicolumn{2}{c}{} \\
  55201.0296 & 0.888 &    32 & 21 & \multicolumn{2}{c}{} &    14 & 23 & \multicolumn{2}{c}{} &    20 & 10 & \multicolumn{2}{c}{} &    43 & 20 & \multicolumn{2}{c}{} \\
  55202.0438 & 0.993 &    20 & 14 & \multicolumn{2}{c}{} &    22 & 16 & \multicolumn{2}{c}{} &     1 &  6 & \multicolumn{2}{c}{} &    24 & 15 & \multicolumn{2}{c}{} \\
\hline\\[-7pt]
\multicolumn{18}{c}{41\,Eri\,A -- more massive, with stronger Hg and Fe lines}\\
\hline\\[-7pt]
  55210.2561      & 0.276 & $-$89 & 22 & 31 & 25              & \multicolumn{2}{c}{} & \multicolumn{2}{c}{} & $-$38 & 23 & \multicolumn{2}{c}{} & \multicolumn{2}{c}{} & \multicolumn{2}{c}{}  \\
  55201.2687      & $^*$0.483 & $-$6  & 21 & \multicolumn{2}{c}{} & \multicolumn{2}{c}{} & \multicolumn{2}{c}{} & $-$3  & 14 & \multicolumn{2}{c}{} & \multicolumn{2}{c}{} & \multicolumn{2}{c}{}  \\ 
  55212.2414      & 0.673 &  36   & 25 & \multicolumn{2}{c}{} & \multicolumn{2}{c}{} & \multicolumn{2}{c}{} &   29  & 10 & \multicolumn{2}{c}{} & \multicolumn{2}{c}{} & \multicolumn{2}{c}{}  \\ 
  55213.2305      & 0.870 & 98    & 47 & \multicolumn{2}{c}{} & \multicolumn{2}{c}{} & \multicolumn{2}{c}{} & 39    & 33 & \multicolumn{2}{c}{} & \multicolumn{2}{c}{} & \multicolumn{2}{c}{}   \\
 \hline\\[-7pt]
\multicolumn{18}{c}{41\,Eri\,B -- less massive  with stronger  Mn and  Ti lines}\\
\hline\\[-7pt]
  55210.2561      & 0.276 & 94     & 35 & \multicolumn{2}{c}{} & \multicolumn{2}{c}{} & \multicolumn{2}{c}{} & 81    & 28 & \multicolumn{2}{c}{} & \multicolumn{2}{c}{} & \multicolumn{2}{c}{} \\ 
  55201.2687     & $^*$0.483 & $-$6   & 21 & \multicolumn{2}{c}{} & \multicolumn{2}{c}{} & \multicolumn{2}{c}{} & $-$3  & 14 & \multicolumn{2}{c}{} & \multicolumn{2}{c}{} & \multicolumn{2}{c}{} \\ 
  55212.2414      & 0.673 & $-$103 & 27 & $-$34 & 30           & \multicolumn{2}{c}{} & \multicolumn{2}{c}{} & $-$31 & 10 & $-$1 & 11            & \multicolumn{2}{c}{} & \multicolumn{2}{c}{} \\
% at 2.5sigma for Fe II (-224+/-90)
  55213.2305      & 0.870 & $-$72  & 29 & \multicolumn{2}{c}{} & \multicolumn{2}{c}{} & \multicolumn{2}{c}{} & $-$40 & 22 & \multicolumn{2}{c}{} & \multicolumn{2}{c}{} & \multicolumn{2}{c}{} \\
\hline\\[-7pt]
\multicolumn{18}{c}{66\,Eri\,A}\\
\hline\\[-7pt]
55203.2457      & 0.063 & $-$17  & 74 & \multicolumn{2}{c}{} & \multicolumn{2}{c}{} & \multicolumn{2}{c}{} & $-$11 & 43 & \multicolumn{2}{c}{} & \multicolumn{2}{c}{} & \multicolumn{2}{c}{}  \\    
55209.2426      & 0.194 &   119  & 86 & \multicolumn{2}{c}{} & \multicolumn{2}{c}{} & \multicolumn{2}{c}{} & 32    & 58 & \multicolumn{2}{c}{} & \multicolumn{2}{c}{} & \multicolumn{2}{c}{}  \\ 
55204.2384      & 0.250 &    27  & 72 & \multicolumn{2}{c}{} & \multicolumn{2}{c}{} & \multicolumn{2}{c}{} & 34    & 26 & \multicolumn{2}{c}{} & \multicolumn{2}{c}{} & \multicolumn{2}{c}{}  \\ 
55210.2476      & 0.384 &     5  & 86 & \multicolumn{2}{c}{} & \multicolumn{2}{c}{} & \multicolumn{2}{c}{} & $-$61 & 54 & \multicolumn{2}{c}{} & \multicolumn{2}{c}{} & \multicolumn{2}{c}{}  \\
55205.2570      & 0.442 &    13  & 74 & \multicolumn{2}{c}{} & \multicolumn{2}{c}{} & \multicolumn{2}{c}{} & $-$1  & 35 & \multicolumn{2}{c}{} & \multicolumn{2}{c}{} & \multicolumn{2}{c}{}  \\
55211.2523       & $^*$0.573 & $-$16  & 47 & \multicolumn{2}{c}{} & \multicolumn{2}{c}{} & \multicolumn{2}{c}{} & $-$20 & 40 & \multicolumn{2}{c}{} & \multicolumn{2}{c}{} & \multicolumn{2}{c}{}  \\
55206.2375      & 0.627 & $-$30  & 66 & \multicolumn{2}{c}{} & \multicolumn{2}{c}{} & \multicolumn{2}{c}{} & $-$40 & 32 & \multicolumn{2}{c}{} & \multicolumn{2}{c}{} & \multicolumn{2}{c}{}  \\
55212.2290      & 0.757 & $-$54  & 49 & \multicolumn{2}{c}{} & \multicolumn{2}{c}{} & \multicolumn{2}{c}{} & $-$71 & 32 & \multicolumn{2}{c}{} & \multicolumn{2}{c}{} & \multicolumn{2}{c}{}  \\
55207.2363      & 0.816 & $-$138 & 62 & \multicolumn{2}{c}{} & \multicolumn{2}{c}{} & \multicolumn{2}{c}{} & $-$25 & 23 & \multicolumn{2}{c}{} & \multicolumn{2}{c}{} & \multicolumn{2}{c}{}  \\
55202.2262      & 0.870 & $-$41  & 73 & \multicolumn{2}{c}{} & \multicolumn{2}{c}{} & \multicolumn{2}{c}{} & $-$48 & 50 & \multicolumn{2}{c}{} & \multicolumn{2}{c}{} & \multicolumn{2}{c}{}  \\ % conjunction phase
 %  55207.2363 & 0.823 &$-$138 & 62 &      &&  & & & & $-$25 &23 & & & & & &  \\
%   55212.2290 & 0.727 & $-$54 & 49 &      &&  & & & & $-$71 &32 &40 &14 & & & &  \\
  %55203.2457 & 0.100   & $-$17 & 74 &    &&  & & & & $-$11 &43 & & & & & &  \\ 
  % 55204.2384 & 0.280   &    27 & 72 &    &&  & & & & 34 &26 & & & & & &  \\ 
  % 55209.2426 & 0.186   &   119 & 86 &    &&  & & & & 32 &58 & & & & & &  \\ 
  % 55205.2570 & 0.465   &    13 & 74 &    &&  & & & &$-$1 &35 & & & & & &  \\
 %  55210.2476 & 0.368   &     5 & 86 &    &&  & & & & $-$61 &54 & & & & & &  \\ 
 %  55206.2375 & 0.642   & $-$30 & 66 &    &&  & & & & $-$40 &32 & & & & & &  \\
  % 55211.2523overl&0.550& $-$16 & 47 &   &&  & & & & $-$20 &40 & & & & & &  \\ % MJDs different
%fom Mak. M values were calculated using
%iraf, are heliocentric and at the middle of the exposure.
%The difference seems to be that they take the start of
%the exposures (200-300 seconds).
\hline\\[-7pt]
\multicolumn{18}{c}{66\,Eri\,B}\\
\hline\\[-7pt]
55203.2445      & 0.063 &    14 & 38 & \multicolumn{2}{c}{} & \multicolumn{2}{c}{} & \multicolumn{2}{c}{} &    76 & 44 & \multicolumn{2}{c}{} & \multicolumn{2}{c}{} & \multicolumn{2}{c}{} \\ 
55209.2415      & 0.194 & $-$95 & 29 & $-$37 & 31           & \multicolumn{2}{c}{} & \multicolumn{2}{c}{} & $-$34 & 42 & \multicolumn{2}{c}{} & \multicolumn{2}{c}{} & \multicolumn{2}{c}{} \\ 
55204.2372      & 0.250 &  $-$7 & 32 & \multicolumn{2}{c}{} & \multicolumn{2}{c}{} & \multicolumn{2}{c}{} & $-$13 & 40 & \multicolumn{2}{c}{} & \multicolumn{2}{c}{} & \multicolumn{2}{c}{} \\ 
55210.2464      & 0.384 & $-$63 & 40 & \multicolumn{2}{c}{} & \multicolumn{2}{c}{} & \multicolumn{2}{c}{} &  $-$6 & 32 & \multicolumn{2}{c}{} & \multicolumn{2}{c}{} & \multicolumn{2}{c}{} \\ 
55205.2558      & 0.442 & $-$81 & 63 & \multicolumn{2}{c}{} & \multicolumn{2}{c}{} & \multicolumn{2}{c}{} & $-$41 & 49 & \multicolumn{2}{c}{} & \multicolumn{2}{c}{} & \multicolumn{2}{c}{} \\
55211.2506      & $^*$0.573 & $-$16 & 47 & \multicolumn{2}{c}{} & \multicolumn{2}{c}{} & \multicolumn{2}{c}{} & $-$20 & 40 & \multicolumn{2}{c}{} & \multicolumn{2}{c}{} & \multicolumn{2}{c}{} \\ 
55206.2364      & 0.627 &    41 & 35 & \multicolumn{2}{c}{} & \multicolumn{2}{c}{} & \multicolumn{2}{c}{} & $-$30 & 37 & \multicolumn{2}{c}{} & \multicolumn{2}{c}{} & \multicolumn{2}{c}{} \\
55212.2273      & 0.757 &    68 & 30 & \multicolumn{2}{c}{} & \multicolumn{2}{c}{} & \multicolumn{2}{c}{} &    61 & 33 & \multicolumn{2}{c}{} & \multicolumn{2}{c}{} & \multicolumn{2}{c}{} \\
55207.2351      & 0.816 &    35 & 23 & \multicolumn{2}{c}{} & \multicolumn{2}{c}{} & \multicolumn{2}{c}{} &    24 & 28 & \multicolumn{2}{c}{} & \multicolumn{2}{c}{} & \multicolumn{2}{c}{} \\
55202.2251      & 0.870 &    16 & 49 & \multicolumn{2}{c}{} & \multicolumn{2}{c}{} & \multicolumn{2}{c}{} &    66 & 43 & \multicolumn{2}{c}{} & \multicolumn{2}{c}{} & \multicolumn{2}{c}{} \\ 
%  55207.2351 & 0.823 &     35 & 23 &   &&  & & & & 24 &28 & & & & & &  \\
%   55212.2273 & 0.727 &     68 & 30 &   &&  & & & &  61 &33 & & & & & &  \\
%  55203.2445 & 0.100 &     14 & 38 &   &&  & & & &  76 &44 & & & & & &  \\ 
%  55204.2372 & 0.280 &   $-$7 & 32 &   &&  & & & & $-$13 &40 & & & & & &  \\ 
%  55209.2415 & 0.186 &  $-$95 & 29 &$-$13&15&  & & & & $-$34 &42 & & & & & &  \\ 
%  55205.2558 & 0.465 &  $-$81 & 63 &   &&  & & & & $-$41 &49 & & & & & &  \\
%  55210.2464 & 0.368 &  $-$63 & 40 &   &&  & & & & $-$6 &32 & & & & & &  \\ 
%  55206.2364 & 0.642 &     41 & 35 &   &&  & & & & $-$30 &37 & & & & & &  \\ 
%  55211.2506overl & 0.550 &$-$16& 47&  &&  & & & & $-$20 &40 & & & & & &  \\ 
\hline
\end{tabular}
%\begin{flushleft}
%Notes:
\tablefoot{
Our phase zero for 66\,Eri refers to conjunction $\mathsc{i}$. Asterisks indicate rotation phases where the spectral lines of both 
components overlap.
% and that of Makaganiuk et al.\ \cite{Makaganiuk2011a} is periastron.
}
%\end{flushleft}
%\end{table}
\end{center}
\end{table*}

\addtocounter{table}{-1}

\begin{table*}
\caption[]{
Continued.
}
%\label{tab:log_meas}
\begin{center}
\begin{tabular}{ccr@{$\pm$}lr@{$\pm$}lr@{$\pm$}lr@{$\pm$}lr@{$\pm$}lr@{$\pm$}lr@{$\pm$}lr@{$\pm$}l}
\hline \hline\\[-7pt]
\multicolumn{1}{c}{MJD} &
\multicolumn{1}{c}{Phase} &
\multicolumn{2}{c}{$\left<B_{\rm z}\right>_{\rm Ti}$} &
\multicolumn{2}{c}{$\left<B_{\rm z}\right>_{\rm Ti,n}$} &
\multicolumn{2}{c}{$\left<B_{\rm z}\right>_{\rm Cr}$} &
\multicolumn{2}{c}{$\left<B_{\rm z}\right>_{\rm Cr,n}$} &
\multicolumn{2}{c}{$\left<B_{\rm z}\right>_{\rm Fe}$} &
\multicolumn{2}{c}{$\left<B_{\rm z}\right>_{\rm Fe,n}$} &
\multicolumn{2}{c}{$\left<B_{\rm z}\right>_{\rm Y}$} &
\multicolumn{2}{c}{$\left<B_{\rm z}\right>_{\rm Y,n}$} \\
&
&
\multicolumn{2}{c}{[G]} &
\multicolumn{2}{c}{[G]} &
\multicolumn{2}{c}{[G]} &
\multicolumn{2}{c}{[G]} &
\multicolumn{2}{c}{[G]} &
\multicolumn{2}{c}{[G]} &
\multicolumn{2}{c}{[G]} &
\multicolumn{2}{c}{[G]} \\
\hline\\[-7pt]
\hline\\[-7pt]
\multicolumn{18}{c}{HD\,33904}\\
\hline\\[-7pt]
  55204.7676 & & $-$32 &11 & \multicolumn{2}{c}{}&\multicolumn{2}{c}{} & \multicolumn{2}{c}{} & $-$3 & 14 & \multicolumn{2}{c}{} & \multicolumn{2}{c}{} & \multicolumn{2}{c}{} \\
\hline\\[-7pt]
\multicolumn{18}{c}{AR\,Aur\,A}\\
\hline\\[-7pt]
  55544.1347 & 0.030 & $-$138 & 97  & \multicolumn{2}{c}{} & \multicolumn{2}{c}{} & \multicolumn{2}{c}{} & $-$160 & 89 & \multicolumn{2}{c}{} & $-$213 & 108 & \multicolumn{2}{c}{} \\
  55553 2003     & 0.223 &  62 & 96  &  \multicolumn{2}{c}{}  & \multicolumn{2}{c}{}  & \multicolumn{2}{c}{} & $-$29    & 76 & \multicolumn{2}{c}{} & $-$101    & 93  & \multicolumn{2}{c}{} \\
  55554.0300     & 0.424 &    266 & 110 & \multicolumn{2}{c}{} & \multicolumn{2}{c}{} & \multicolumn{2}{c}{} & 82     & 41 & \multicolumn{2}{c}{} & 254    & 88  & \multicolumn{2}{c}{} \\ 
 % 55195.1390 & 0.623pr &  $-$462 &108  & 10 & 10 &  & & & & $-$289 &73 &10 &10 &$-$452 &146 &10 &10  \\ 
  55555.1200     & 0.687 & $-$199 & 63  & 64    & 65           & \multicolumn{2}{c}{} & \multicolumn{2}{c}{} & $-$207 & 67 & 33    & 69           & $-$216 & 68  & 78    & 72  \\ 
  55556.0350     & 0.908 & $-$335 & 94  & $-$70 & 98           & \multicolumn{2}{c}{} & \multicolumn{2}{c}{} & $-$220 & 59 & $-$30 & 65           & $-$353 & 100 & $-$50 & 112  \\ 
  \hline\\[-7pt]
\multicolumn{18}{c}{AR\,Aur\,B}\\
\hline\\[-7pt]
  55544.1347 & 0.030 & 208    & 116 & \multicolumn{2}{c}{} & \multicolumn{2}{c}{} & \multicolumn{2}{c}{} & 263    & 85 & $-$61 & 80           & \multicolumn{2}{c}{} & \multicolumn{2}{c}{}  \\
  55553 2003     & 0.223 & 201    & 136 & \multicolumn{2}{c}{} & \multicolumn{2}{c}{} & \multicolumn{2}{c}{} & 140     & 58 & \multicolumn{2}{c}{} & \multicolumn{2}{c}{} & \multicolumn{2}{c}{}  \\
  55554.0300     & 0.424 & $-$162 & 61  & \multicolumn{2}{c}{} & \multicolumn{2}{c}{} & \multicolumn{2}{c}{} & $-$116 & 63 & \multicolumn{2}{c}{} & \multicolumn{2}{c}{} & \multicolumn{2}{c}{}  \\ 
 % 55195.1390 & 0.623pr &  $-$448 &318  & \multicolumn{2}{c}{} & & & & & 392 &54 &10 &10 & & & &  \\ 
  55555.1200     & 0.687 & 196    & 73  & \multicolumn{2}{c}{} & \multicolumn{2}{c}{} & \multicolumn{2}{c}{} & 165     & 74 & \multicolumn{2}{c}{} & \multicolumn{2}{c}{} & \multicolumn{2}{c}{}  \\ 
  55556.0350     & 0.908 & 71     & 102 & \multicolumn{2}{c}{} & \multicolumn{2}{c}{} & \multicolumn{2}{c}{} & 93     & 89 & \multicolumn{2}{c}{} & \multicolumn{2}{c}{} & \multicolumn{2}{c}{}  \\ 
  \hline\\[-7pt]
\multicolumn{18}{c}{HD\,53244}\\
\hline\\[-7pt]
  55204.2799 & & 62 & 43 & \multicolumn{2}{c}{} & 85 & 49 & \multicolumn{2}{c}{} & 98 & 46 & \multicolumn{2}{c}{} & \multicolumn{2}{c}{} & \multicolumn{2}{c}{} \\
\hline\\[-7pt]
\multicolumn{18}{c}{HD\,78316}\\
\hline\\[-7pt]
  55202.3038 & & $-$22 & 15 & \multicolumn{2}{c}{}  & $-$52 & 23 & \multicolumn{2}{c}{} & $-$32 & 10 & $-$4 & 11            & \multicolumn{2}{c}{} & \multicolumn{2}{c}{} \\
  55211.2873 & & 84    & 38 & \multicolumn{2}{c}{} & $-$13 & 18 & \multicolumn{2}{c}{} & 3     & 17 & \multicolumn{2}{c}{} & \multicolumn{2}{c}{} & \multicolumn{2}{c}{} \\
  \hline\\[-7pt]
\multicolumn{18}{c}{HD\,101189}\\
\hline\\[-7pt]
%54982.4744 & & \multicolumn{2}{c}{} & \multicolumn{2}{c}{} & $-$340 & 48 & 121 & 100?            & $-$237 & 449 & \multicolumn{2}{c}{} & $-$138 & 232 & \multicolumn{2}{c}{} \\
  55201.3629 & & $-$74 & 24           & 22 & 25              & $-$24  & 18 & \multicolumn{2}{c}{} & $-$42  & 26  & \multicolumn{2}{c}{} & $-$37  & 11  & $-$6 & 12 \\
  \hline\\[-7pt]
\multicolumn{18}{c}{HD\,179761}\\
\hline\\[-7pt]
  55319.4267& & \multicolumn{2}{c}{} & \multicolumn{2}{c}{} & \multicolumn{2}{c}{} & \multicolumn{2}{c}{} & $-$88 & 49 & \multicolumn{2}{c}{} & \multicolumn{2}{c}{} & \multicolumn{2}{c}{} \\
\hline
\multicolumn{18}{c}{HD\,209459}\\
\hline\\[-7pt]
  %55417.1053&  &195  &82  & && 158&51 &10 &10 & 36 &24 & & & & & &  \\
  55417 1494 & & 99 & 22 & 24 & 21              & 68 & 20 & $-$5 & 19            & 53 & 17 & 9    & 18 & \multicolumn{2}{c}{} & \multicolumn{2}{c}{}  \\
  55421.2598 & & 31 & 16 & \multicolumn{2}{c}{} & 12 & 19 & \multicolumn{2}{c}{} & 27 & 10 &\multicolumn{2}{c}{}  & \multicolumn{2}{c}{} & \multicolumn{2}{c}{}  \\ 
  \hline\\[-7pt]
\multicolumn{18}{c}{HD\,221507}\\
\hline\\[-7pt]
  55210.0485& & 78 & 44 & \multicolumn{2}{c}{} & 95 & 58 & \multicolumn{2}{c}{} & 56 & 24 & \multicolumn{2}{c}{} & 78 & 25 & $-$14 & 24 \\
\hline
\end{tabular}
\end{center}
\end{table*}

The HARPS archive spectra cover the wavelength range 3780--6913\,\AA{}, with a small gap around 5300\,\AA{}. 
Each observation of the star is usually split into four to eight sub-exposures, 
obtained with four different orientations of the quarter-wave retarder plate relative to the 
beam splitter of the circular polarimeter. The reduction was performed using the HARPS data reduction 
software available at the ESO headquarters in Germany. 
The Stokes~$I$ and $V$ parameters were derived following the ratio method described by 
Donati et al.\ (\cite{Donati1997}), 
ensuring in particular that all spurious signatures are removed at first order. 
%and the diagnostic null spectrum were deduced with the help of the “ratio method” described by Bagnulo et al.\ (\cite{Bagnulo2009}).
Null polarisation spectra (labeled with $n$ in Table~\ref{tab:log_meas}) were calculated by combining the sub-exposures 
in such a way that the polarisation cancels out, allowing us to verify that no spurious signals are present in the data.
Other details of the typical acquisition and calibration of the HARPSpol observations can be found in 
papers by Makaganiuk et al.\ (e.g. \cite{Makaganiuk2011a}, \cite{Makaganiuk2011b}). 

\begin{table}
\caption[]{Line lists for the different elements used in the magnetic field measurements.
% of mean longitudinal and mean quadratic magnetic fields.
}
\begin{center}
\begin{tabular}{rrcrr}
\hline
\hline
\multicolumn{1}{c}{$\lambda$ [\AA{}] \rule{0pt}{2.4ex}} &
\multicolumn{1}{c}{$g_{\rm eff}$} &
&
\multicolumn{1}{c}{$\lambda$ [\AA{}]} &
\multicolumn{1}{c}{$g_{\rm eff}$} \\
\hline
\multicolumn{2}{c}{\ion{Ti}{ii}} & & \multicolumn{2}{c}{\ion{Fe}{ii}} \\
\hline
4012.385 & 0.716 & & 4024.547 & 1.095 \\
4163.648 & 1.071 & & 4122.668 & 1.326 \\ 
4290.219 & 1.092 & & 4178.862 & 0.924 \\ 
4294.099 & 1.203 & & 4273.326 & 2.155 \\ 
4300.049 & 1.205 & & 4296.572 & 0.581 \\ 
4386.844 & 0.929 & & 4385.387 & 1.330 \\
4394.051 & 1.335 & & 4416.830 & 0.767 \\ 
4395.033 & 1.074 & & 4515.339 & 1.044 \\ 
4399.772 & 1.400 & & 4520.224 & 1.337 \\ 
4409.516 & 1.457 & & 4583.837 & 1.144 \\ 
4411.074 & 0.900 & & 4596.015 & 1.591 \\ 
4417.719 & 0.795 & & 4620.521 & 1.305 \\
4443.794 & 0.923 & & 4629.339 & 1.314 \\ 
4468.507 & 1.048 & & 4635.316 & 1.042 \\
4488.331 & 1.072 & & 4731.453 & 0.655 \\ 
4501.273 & 0.913 & & 4923.927 & 1.695 \\ 
4563.761 & 0.985 & & 5001.959 & 1.142 \\
4571.968 & 0.944 & & 5197.577 & 0.671 \\
4805.085 & 1.140 & & 5234.625 & 0.869 \\ 
4911.193 & 1.865 & & 5260.259 & 1.168 \\
 & & & 5506.195&1.150 \\
 & & & 6416.919&1.460 \\
 & & & 6456.383&1.182 \\
%Ti II: 4409.516 (1.457), 4468.507  (1.048), 4488.331 (1.072),   
%Fe II: 4024.547 (1.095), 4620.521 (1.305),  
\hline
\multicolumn{2}{c}{\ion{Cr}{ii}} & & \multicolumn{2}{c}{\ion{Y}{ii}} \\
\hline
3952.594 & 1.035 & & 3950.352 & 1.115 \\ 
4177.529 & 0.880 & & 4054.076 & 1.503 \\
4198.277 & 1.985 & & 4269.277 & 0.798 \\
4242.364 & 1.183 & & 4284.188 & 0.521 \\
4261.913 & 1.084 & & 4309.631 & 1.170 \\ 
4275.567 & 0.922 & & 4358.728 & 0.995 \\ 
4588.199 & 1.059 & & 4374.935 & 0.955 \\
4589.901 & 2.102 & & 4398.013 & 0.995 \\
4592.049 & 1.201 & & 4422.591 & 0.500 \\
4616.629 & 0.793 & & 4554.988 & 1.332 \\
4618.803 & 0.914 & & 4682.324 & 1.245 \\
4848.235 & 1.247 & & 4883.684 & 1.155 \\
5123.211 & 0.550 & & 4900.120 & 1.020 \\
5237.329 & 1.335 & & 5087.416 & 1.255 \\
5478.365 & 1.136 & & 5200.406 & 0.715 \\
         &       & & 5205.724 & 1.085 \\
         &       & & 5402.774 & 0.900 \\
         &       & & 5509.895 & 0.790 \\
         &       & & 5662.925 & 1.000 \\
%YII 3952.594 (1.035), 4177.529 (0.880), 4198.277 (1.985), 5123.211 (0.550), 
%Cr II 4054.076 (1.503), 4284.188 (0.521), 4554.988 (1.332), 4269.277 (0.798) 
\hline
\end{tabular}
\end{center}
\label{tab:lines}
\end{table} 
 
The diagnostic potential of high-resolution circularly polarised spectra using the moment technique has been 
discussed at length in numerous papers by Mathys (e.g. \cite{Mathys1993}, \cite{Mathys1995a}, \cite{Mathys1995b}).
Wavelength shifts between right- and left-hand side circularly polarised spectra are
interpreted in terms of a longitudinal magnetic field $\left<B_{\rm z}\right>$.
%using the moment technique. described by Mathys (\cite{Mathys1994}).
The major problem in the analysis of high-resolution spectra is the proper line identification
of blend-free spectral lines. The quality of the selection varies strongly from star to star,
depending on the binarity, the line broadening, and the richness of the spectrum. In addition, due
to the inhomogeneous element distribution on the stellar surface, the degree of blending
changes with the stellar rotation phase. All the lines we tried to employ in the diagnosis
of the magnetic fields on the surface of our target stars are presented in Table~\ref{tab:lines},
together with their Land\'e factors. We note that the actual list of lines measured in each star
on each spectrum can differ from one observation to the next due to the different quality 
of the spectra and variable blending over the rotation cycle. For the lines of iron-peak elements,
the Land\'e factors were 
taken from Kurucz's  list of atomic data (Kurucz \cite{Kurucz1989}), while the  Land\'e factors
for the Y lines were retrieved from the Vienna Atomic Line Database (VALD; e.g.\ Kupka et al.\ \cite{Kupka1999}). 

The measurements of the mean longitudinal magnetic field in our targets using line lists of different 
elements separately are presented in Table~\ref{tab:log_meas}. Since the element abundances are different 
in different stars, only the lines of ions with a larger number of measurable lines were used for the analysis.
A study of elements with a small number of lines sets considerable limitations on the amount of 
information on the presence of a magnetic field. The most populated samples of lines belong to Ti and Fe,
followed by Y and Cr. In a number of stars, Cr and Y lines appear too faint and could not be used
in the measurements. 
The mean longitudinal magnetic fields
were determined from null spectra only in the phases where the longitudinal field was detected at a 3$\sigma$ 
significance level. Since no significant fields could be determined from null spectra, we conclude that
any noticeable spurious polarisation is absent. The measurements on the spectral lines of individual elements
using null spectra are labeled with $n$ in Table~\ref{tab:log_meas}.

To calculate the rotation phases for the star HD\,11753, the following ephemeris was used: 

\begin{eqnarray}
 {\rm T_{max}} = 2451800.0 + (9.531 \pm 0.001) E  \nonumber,
%{\rm HD\,11753}: {\rm HJD}2451800.0 + 9.531 \pm 0.001 E  \nonumber
%{\rm HD\,11753}: {\rm HJD}2451800.0 &\pm& 0. + 9.531 \pm 0.001 E  \nonumber
%{\rm HD\,11753}: {\rm HJD}2451800.0 &\pm& 0. + 9.531 \pm 0.001 E  \nonumber \\
%%\nonumber \\
%%T(Weq=max) = 2455000.672
%{\rm  HD\,150193}: \left<B_{\rm z}\right>^{\rm pos~ext} &=& \nonumber \\
%{\rm MJD}55318.05538 &\pm& 0.03072 + 1.31697 \pm 0.00013 E  \nonumber \\
%%\nonumber \\
\end{eqnarray}
\noindent
where  T$_{max}$ is the HJD for which the equivalent width of \ion{Y}{ii} is at maximum.

This ephemeris is based on the work by Briquet et al.\ (\cite{Briquet2010}) and Korhonen et al.\ (2012, in preparation), 
who studied numerous 
spectroscopic time series of this star acquired in the last decade.
The discussion on the rotation periods of the
three SB2 systems, 41\,Eri, 66\,Eri, and AR\,Aur is presented in the next section, where we conclude
that the binary components in all systems apart from the  66\,Eri system are already synchronised and their rotation periods 
correspond to their orbital period.
The adopted ephemeris to calculate rotation phases for
41\,Eri and 66\,Eri is discussed in detail in the same section, while the rotation phases for AR\,Aur
were calculated using ephemeris presented in the work of Albayrak et al.\ (\cite{Albayrak2003}).

Another approach to study the presence of magnetic fields in upper main sequence 
stars is to determine the value of the mean quadratic 
magnetic field,
\begin{displaymath}
\langle B_q\rangle= (\langle B^2\rangle + \langle B_z^2\rangle)^{1/2},
\end{displaymath}
which is derived through the application of the moment technique, 
described, e.g.\ by Mathys \& Hubrig (\cite{Mathys2006}). Here, $\langle B^2\rangle$ is 
the mean square magnetic field modulus, i.e.\ the average over the stellar 
disc of the square of the modulus of the magnetic field vector, weighted by 
the local emergent line intensity, while $\langle B_z^2\rangle$ is 
the mean square longitudinal magnetic field, i.e.\ the average over the stellar 
disc of the square of the line-of-sight component of the magnetic 
vector, weighted by the local emergent line intensity. 

The mean quadratic magnetic field is determined from the study of the 
second-order moments of the line profiles recorded in unpolarised 
light (that is, in the Stokes parameter $I$). 
The second-order moment $R_I^{(2)}(\lambda_I) $ of a spectral 
line profile recorded in unpolarised light about its center of 
gravity $\lambda_I$ is defined as

\begin{displaymath}
R_I^{(2)}(\lambda_0) ={1\over W_\lambda}~\int{r_{{\cal
F}_I}(\lambda-\lambda_0)(\lambda-\lambda_0)^2}~{\rm d}\lambda .
\end{displaymath}

\begin{table*}
\caption[]{
Measurements of the quadratic magnetic field using sets of lines belonging to different elements.
All quoted errors are 1\,$\sigma$ uncertainties.
%The FORS\,1 measurements  
%published by Hubrig et al.\ (\cite{Hubrig2009}) are presented in the first line, respectively.
%Logbook of the spectropolarimetric observations of , and of the
%magnetic field measurements.
%Results of our magnetic field measurements.
}
\label{tab:log_meas2}
\begin{center}
\begin{tabular}{ccr@{$\pm$}lr@{$\pm$}lr@{$\pm$}lr@{$\pm$}l}
\hline \hline\\[-7pt]
\multicolumn{1}{c}{MJD} &
\multicolumn{1}{c}{Phase} &
\multicolumn{2}{c}{$\left<B_{\rm q}\right>_{\rm Ti}$} &
%\multicolumn{2}{c}{$\left<B_{\rm q}\right>_{\rm Ti,n}$} &
\multicolumn{2}{c}{$\left<B_{\rm q}\right>_{\rm Cr}$} &
%\multicolumn{2}{c}{$\left<B_{\rm q}\right>_{\rm Cr,n}$} &
\multicolumn{2}{c}{$\left<B_{\rm q}\right>_{\rm Fe}$} &
%\multicolumn{2}{c}{$\left<B_{\rm q}\right>_{\rm Fe,n}$} &
\multicolumn{2}{c}{$\left<B_{\rm q}\right>_{\rm Y}$} \\
%\multicolumn{2}{c}{$\left<B_{\rm q}\right>_{\rm Y,n}$} \\
&
&
\multicolumn{2}{c}{[G]} &
\multicolumn{2}{c}{[G]} &
\multicolumn{2}{c}{[G]} &
\multicolumn{2}{c}{[G]} \\
%\multicolumn{2}{c}{[G]} &
%\multicolumn{2}{c}{[G]} &
%\multicolumn{2}{c}{[G]} &
%\multicolumn{2}{c}{[G]} \\
\hline\\[-7pt]
\multicolumn{10}{c}{HD\,11753} \\
\hline\\[-7pt]
  55206.0271 & 0.413 & 2034 & 669 & \multicolumn{2}{c}{} & \multicolumn{2}{c}{} & \multicolumn{2}{c}{} \\
  55202.0438 & 0.993 & 2168 & 502 & \multicolumn{2}{c}{} & \multicolumn{2}{c}{} & \multicolumn{2}{c}{} \\
\hline\\[-7pt]
\multicolumn{10}{c}{41\,Eri\,A -- more massive, with stronger Hg and Fe lines}\\
\hline\\[-7pt]
 55213.2305 & 0.870 & 4520 & 758 & \multicolumn{2}{c}{} & \multicolumn{2}{c}{} & \multicolumn{2}{c}{} \\
\hline\\[-7pt]
\multicolumn{10}{c}{41\,Eri\,B -- less massive, with stronger Mn and Ti lines}\\
\hline\\[-7pt]
  55212.2414 & 0.673 & 3885 & 1115 & \multicolumn{2}{c}{} & 2332 & 488 & \multicolumn{2}{c}{} \\ 
  55213.2305 & 0.870 & 4484 & 1201 & \multicolumn{2}{c}{} & 2106 & 634 & \multicolumn{2}{c}{} \\
\hline\\[-7pt]
\multicolumn{10}{c}{66\,Eri\,A}\\
\hline\\[-7pt]
55203.2457 & 0.100 & \multicolumn{2}{c}{} & \multicolumn{2}{c}{} & 3955 & 1216          & \multicolumn{2}{c}{} \\   
55205.2570 & 0.465 & \multicolumn{2}{c}{} & \multicolumn{2}{c}{} & 4941 & 1174          & \multicolumn{2}{c}{} \\
55206.2375 & 0.642 & 7687 & 2293          & \multicolumn{2}{c}{} & 5861 & 1048          & \multicolumn{2}{c}{} \\
55212.2290 & 0.727 & 7915 & 1760          & \multicolumn{2}{c}{} & \multicolumn{2}{c}{} & \multicolumn{2}{c}{} \\
55202.2262 & 0.916 & \multicolumn{2}{c}{} & \multicolumn{2}{c}{} & 5899 & 1940          & \multicolumn{2}{c}{} \\
\hline\\[-7pt]
\multicolumn{10}{c}{66\,Eri\,B}\\
\hline\\[-7pt]
55203.2457 & 0.100 & 5480 & 1175          & \multicolumn{2}{c}{} & \multicolumn{2}{c}{} & \multicolumn{2}{c}{} \\   
55209.2426 & 0.186 & 4306 & 805           & \multicolumn{2}{c}{} & \multicolumn{2}{c}{} & \multicolumn{2}{c}{} \\
55210.2476 & 0.368 & 4070 & 1239          & \multicolumn{2}{c}{} & \multicolumn{2}{c}{} & \multicolumn{2}{c}{} \\
55205.2570 & 0.465 & \multicolumn{2}{c}{} & \multicolumn{2}{c}{} & 7271 & 1873          & \multicolumn{2}{c}{} \\
55207.2362 & 0.823 & 4980 & 931           & \multicolumn{2}{c}{} & \multicolumn{2}{c}{} & \multicolumn{2}{c}{} \\
55202.2262 & 0.916 & 3834 & 1247          & \multicolumn{2}{c}{} & \multicolumn{2}{c}{} & \multicolumn{2}{c}{} \\
\hline\\[-7pt]
\multicolumn{10}{c}{HD\,33904}\\
\hline
55204.7676& & 4835 & 1402 & \multicolumn{2}{c}{} & \multicolumn{2}{c}{} & \multicolumn{2}{c}{} \\
\hline
\multicolumn{10}{c}{AR\,AurA}\\
\hline\\[-7pt]
55553 2003 & 0.223 & 6339 & 1879 & \multicolumn{2}{c}{} & \multicolumn{2}{c}{} & \multicolumn{2}{c}{} \\
55555.1200 & 0.687 & 6721 & 2231 & \multicolumn{2}{c}{} & \multicolumn{2}{c}{} & \multicolumn{2}{c}{} \\
55556.0350 & 0.908 & 5849 & 1550 & \multicolumn{2}{c}{} & \multicolumn{2}{c}{} & 8238 & 1750 \\
\hline
\multicolumn{10}{c}{HD\,101189}\\
\hline\\[-7pt]
  55201.3629 & & 5696 & 709 & 2731 & 842 & 3420 & 1041 & \multicolumn{2}{c}{} \\
\hline\\[-7pt]
\multicolumn{10}{c}{HD\,179761}\\
\hline\\[-7pt]
  55204.2799 & & \multicolumn{2}{c}{} & \multicolumn{2}{c}{} & 3246 & 1052 & \multicolumn{2}{c}{} \\
\hline
\multicolumn{10}{c}{HD\,209459}\\
\hline\\[-7pt]
55417 1494 & & \multicolumn{2}{c}{} & 1269 & 422 & \multicolumn{2}{c}{} & \multicolumn{2}{c}{} \\  
55421.2598 & & \multicolumn{2}{c}{} & 1470 & 409 & \multicolumn{2}{c}{} & \multicolumn{2}{c}{} \\
\hline
\end{tabular} 
\end{center}
\end{table*}

The integration runs over the whole width of the observed 
line (see Mathys \cite{Mathys1988} for details). ${W_\lambda}$ is the 
line equivalent width; $r_{{\cal F}_I} $ is the line profile
\begin{displaymath}
r_{{\cal F}_I}
=1-({\cal F}_I/{\cal F}_{I_{\rm c}}) .
\end{displaymath}
 $\ensuremath{{\cal F}_I} $ (resp. $\ensuremath{{\cal F}_{I_{\rm c}}} $)
is the integral over the visible stellar disk of the emergent intensity
in the line (resp. in the neighboring continuum).
The analysis is usually based on the consideration of samples 
of reasonably unblended lines and critically depends on the
%size of the set
number of lines that can be employed. 
Importantly, contrary to the 
mean longitudinal field, the mean quadratic magnetic field provides a measurement of the field strength 
that is fairly
insensitive to its structure. Thus it is especially well suited to detect  
fields that have a complex structure, as is likely the case in 
HgMn stars. Using this method, 
Mathys \& Hubrig (\cite{MathysHubrig1995}) could demonstrate the presence of quadratic 
magnetic fields in two close double-lined systems with HgMn primary 
stars, 74 Aqr and $\chi$ Lup.
The measurements of the mean quadratic magnetic field in our targets are presented in Table~\ref{tab:log_meas2}.  

The crossover effect can be measured by the second-order moment about the centre of the profiles of 
spectral lines recorded in the Stokes parameter $V$.  
Mathys (\cite{Mathys1995a}) showed that one can derive from the measurements a quantity called the 
mean asymmetry of the longitudinal magnetic field, which is the first moment of the component of the 
magnetic field along the line of sight, about the plane defined by the line of sight, and the stellar 
rotation axis. The secon- order moment of a line profile recorded in the Stokes parameter $V$ with respect to
the wavelength $\lambda_0$ of the centre of gravity of the corresponding unpolarised profile is defined as

\begin{eqnarray}
R_V^{(2)}(\lambda_0) &=& \frac{1}{W_{\lambda}} \int r_{{\cal F}_V} (\lambda-\lambda_0) (\lambda-\lambda_0)^2 d\lambda, \nonumber
\end{eqnarray}

\noindent
where $W_{\lambda}$ is the equivalent width of the line and $r_{{\cal F}_V}$ is the line profile in the 
Stokes parameter $V$.
A detectable crossover effect was found only in very few stars. These measurements are discussed 
in the following section.

\subsection{SOFIN observations of AR\,Aur}

Spectropolarimetric observations of the double-lined eclipsing binary AR\,Aur were obtained on 
five nights between December 12 and December 26, 2010. 
%with S/N$>$250
We used the low-resolution camera
%($R=\lambda/\Delta\lambda\approx30,000$)
($R\approx30\,000$)
of the echelle spectrograph SOFIN (Tuominen et al.\ \cite{Tuominen1999}),
mounted at the Cassegrain focus of the Nordic Optical Telescope (NOT).
With the 2K Loral CCD detector, we registered 40 echelle orders partially covering
the range from 3500 to 10\,000\,\AA{}, 
with a length of the spectral orders of about 140\,\AA{} at 5500\,\AA{}.
The polarimeter is located in front of the entrance slit of the spectrograph and consists of a fixed 
calcite beam splitter aligned along the slit and a rotating super-achromatic quarter-wave plate. Two spectra 
polarised in opposite sense are recorded simultaneously for each echelle order, providing sufficient separation 
by the cross-dispersion prism.
Two to four sub-exposures with the quarter-wave plate angles separated 
by $90^\circ$ were used to derive circularly polarised spectra. A detailed description of the 
SOFIN spectropolarimeter and its polarimetric data reduction is given in Ilyin (\cite{Ilyin2012}).
The spectra are reduced with the 4A software package (Ilyin \cite{Ilyin2000}). 
Bias subtraction, master flat-field correction, 
scattered light subtraction, and weighted extraction of spectral orders comprise the standard steps 
of the image processing. A ThAr spectral lamp is used for wavelength calibration, taken before
and after each target exposure to minimise temporal variations in the spectrograph. 

The results of the longitudinal and quadratic magnetic field measurements using line lists for Ti and Fe  
are presented in Tables~\ref{tab:log_meas} and \ref{tab:log_meas2}.
Due to the lower spectral resolution of our SOFIN spectra and the somewhat lower S/N achieved with a smaller 
diameter telescope, the measurement uncertainties for AR\,Aur are on average larger than those achieved in 
the HARPS spectra. In addition, the line widths of the order of 23\,km\,s$^{-1}$ in 
this system exceed the line widths for most of the targets in our sample.

\subsection{FORS\,1/2 measurements of the PGa star HD\,19400 and the HgMn star HD\,65949}

A few polarimetric spectra of the PGa star HD\,19400 and the HgMn star HD\,65949 were previously 
obtained with 
FORS\,1 (Hubrig et al.\ \cite{Hubrig2006b}, \cite{Hubrig2011})
and most recently with FORS\,2\footnote{
The spectropolarimetric capabilities of FORS\,1 were moved to
FORS\,2 in 2009.
}
on Antu (UT1) from May 2011 to January 2012.
The star HD\,19400 belongs to the group of PGa stars, presenting the hotter extension
of the HgMn stars. PGa stars exhibit strongly overabundant P and Ga and deficient He.
Maitzen (\cite{Maitzen1984}) suggested the presence of a magnetic field in HD\,19400, using observations of 
the $\lambda$5200 feature. 
Similar to the group of HgMn stars, the spectra of HD\,19400 reveal a strong
overabundance of the elements Mn and Hg (Alonso et al.\ \cite{Alonso2003}).
This star is likely a binary system; Dommanget \& Nys (\cite{DommangetNys2002}) mention in the 
CCDM catalogue a nearby component at a separation of 0\farcs1 and a position angle of 179$^{\circ}$.

The spectrum of the  HgMn star HD\,65949 was studied in detail by  Cowley et al.\ (\cite{Cowley2010}),
who discovered that this star exhibits 
enormous enhancements of the elements rhenium through mercury ($Z = 75-80$). 
In the catalogue of HgMn stars (Schneider \cite{Schneider1981}), HD\,65949 is mentioned as an SB1 system.

During the observations with FORS\,2, we used a slit width of 0\farcs4 and the GRISM 600B to achieve a 
spectral resolving power of about 2000.
 A detailed description of the assessment of the longitudinal 
magnetic field measurements using FORS\,2 is presented in our previous work
(e.g. Hubrig et al.\ \cite{Hubrig2004a, Hubrig2004b}, and references therein). 
%We repeat here the major steps of the magnetic field determination. 
The mean longitudinal 
magnetic field, $\left< B_{\rm z}\right>$, was derived using 

\begin{eqnarray} 
\frac{V}{I} = -\frac{g_{\rm eff} e \lambda^2}{4\pi{}m_ec^2}\ \frac{1}{I}\ 
\frac{{\rm d}I}{{\rm d}\lambda} \left<B_{\rm z}\right>,  \nonumber
%\label{eqn:one}
\end{eqnarray} 

\noindent 
where $V$ is the Stokes parameter that measures the circular polarisation, $I$ 
is the intensity in the unpolarised spectrum, $g_{\rm eff}$ is the effective 
Land\'e factor, $e$ is the electron charge, $\lambda$ is the wavelength, $m_e$ the 
electron mass, $c$ the speed of light, ${{\rm d}I/{\rm d}\lambda}$ is the 
derivative of Stokes~$I$, and $\left<B_{\rm z}\right>$ is the mean longitudinal magnetic 
field. 

\begin{table}
\caption[]{
Magnetic field measurements of HD\,19400 and HD\,65949 using FORS\,1/2. Already published measurements
are marked by asterisks.
All quoted errors are 1$\sigma$ uncertainties.
}
\label{tab:log_measfors}
\centering
%\begin{tabular}{ccr@{$\pm$}lr@{$\pm$}l}
\begin{tabular}{rr@{$\pm$}lr@{$\pm$}l}
\hline \hline\\[-7pt]
\multicolumn{1}{c}{MJD} &
%\multicolumn{1}{c}{Phase} &
\multicolumn{2}{c}{$\left<B_{\rm z}\right>_{\rm all}$ [G]} &
\multicolumn{2}{c}{$\left<B_{\rm z}\right>_{\rm hyd}$ [G]} \\
\hline\\[-7pt]
\multicolumn{5}{c}{HD\,19400} \\
\hline\\[-7pt]
  $^{*}$52852.371 &   151 & 46 &   217 & 65 \\
  55845.295       &    14 & 24 &    32 & 26 \\
  55935.109       & $-$65 & 26 & $-$110& 30 \\
\hline\\[-7pt]
\multicolumn{5}{c}{HD\,65949}\\
\hline\\[-7pt]
  $^{*}$53002.082 &$-$290  &62  & $-$143 & 71 \\
  $^{*}$54108.177 & $-$116 & 32 & $-$131 & 39 \\
  $^{*}$54433.366 & $-$18  & 21 &  $-$18 & 30 \\
  55686.070 &  $-$77  & 24 &  $-$63 & 24 \\
  55688.040 &  $-$182 & 34 & $-$163 & 34 \\
\hline
\end{tabular}
\end{table}

The mean longitudinal magnetic field was measured either by using only the absorption hydrogen Balmer
lines or by using the entire spectrum, including all available absorption lines. Our measurements, 
acquired over a few years, are presented
in Table~\ref{tab:log_measfors}, together with the modified Julian dates of mid-exposure. A few 
weak field detections 
at a significance level higher than $3\sigma$ were achieved for each star.
As the rotation periods for both stars are unknown, it is not possible at the present stage to  conclude whether the 
variability of the magnetic field is caused by a rotational modulation.

With respect to the reliability of our FORS\,2 measurements of rather weak longitudinal 
magnetic fields, we note that
the feasibility of such measurements in stars of different mass and
at different evolutionary stages
using FORS\,1/2 in spectropolarimetric mode has been demonstrated by numerous studies during the last
ten years. It has also been confirmed by comparing measurements of the magnetic field for
well-studied magnetic stars with values from the literature.
These stars have typically strong longitudinal magnetic fields of the order of a few kG.
What is not yet so well established are the measurements of several calibrators with magnetic fields well below 1000\,G.
The FORS $V$/$I$ spectrum is deduced from four spectra:
the two spectra for the ordinary and extraordinary beams coming out of the
Wollaston prism, taken at two different settings of the retarder wave plate.
They are combined using the following equation:

\begin{eqnarray}
\frac{V}{I} =
\frac{1}{2} \left\{ \left( \frac{f^{\rm o} - f^{\rm e}}{f^{\rm o} + f^{\rm e}} \right)_{\alpha=-45^{\circ}}
- \left( \frac{f^{\rm o} - f^{\rm e}}{f^{\rm o} + f^{\rm e}} \right)_{\alpha=+45^{\circ}} \right\}, \nonumber
%\label{eqn:two}  
\end{eqnarray}

While $f^o_{-45}$ and $f^e_{-45}$, as well as $f^o_{+45}$ and $f^e_{+45}$
are taken at the same time, respectively, this is obviously not true for the
two pairs and leads to differences in flux levels.
Also, the transmission for the ordinary and extraordinary beams is not the
same.
In the equation used for the determination of the $V$/$I$ spectrum above,
both varying transmission and flux levels at different
times are compensated through the double difference, especially removing
the need for a flatfield.
Still, there might be residuals left after using the above equation that come from
instrumental effects, e.g.\ crosstalk from linear to circular polarisation.
%One way to calculate the FORS\,2 null spectrum, which should not show any signature of a magnetic field,
%would be to change the sign in the
%equation above from ``$-$'' to ``+'', leaving only the error terms.
%This does not work well for FORS\,1/2, since the different flux levels do not cancel.
%The approach promoted by Donati et al.\ (\cite{Donati1997}), which
%calculates the $V$ and $N$
%spectra rather from the relations of the four (or eight) spectra from the
%differences, also suffers from these unbalances.
%Even if a more sophisticated treatment is employed, it is clear that e.g.\ the
%standard quality of the flatfields (frequently just one frame is provided) is not  sufficient to calibrate
%the very high $S/N$ spectra used for the determination of the magnetic field strengths.
The null spectrum is usually determined as the difference between two consecutive $V$/$I$ spectra.
This computation should get rid of the magnetic field signatures and leave only the instrumental
features.
Yet, different signal-to-noise ratios in the consecutive $V$/$I$ spectra might introduce spurious
null signals.

On the other hand, a few recent FORS\,2 spectropolarimetric observations over the rotation/magnetic period of the
bright classical Ap star HD\,142070, with a weak magnetic field, typical for a number
of objects observed with FORS\,1/2, revealed a good agreement with the measurements obtained with a
high-resolution spectropolarimeter (Mathys et al.\ \cite{Mathys2012}). These results indicate that the 
order of magnitude of the uncertainties achieved with FORS\,2 is correct.

\section{Results for individual targets}
\label{sect:indivi}

{\it HD\,11753:}

Among the studied targets, the star HD\,11753 with eleven HARPS observations has the best  
rotation phase coverage.  
This system is an SB1 with a period of 41.489\,d,
according to the 9$^{\rm th}$ Catalogue of Spectroscopic Binary Orbits (Pourbaix et al.\ \cite{Pourbaix2009}).
It is also an astrometric Hipparcos binary according to the CCDM catalogue 
(Dommanget \& Nys \cite{DommangetNys2002}).

The spectra of HD\,11753 exhibit a pronounced variability of Ti, Cr, Sr, and Y, and their surface 
inhomogeneous distribution was studied in detail by Briquet et al.\ (\cite{Briquet2010}) and Makaganiuk
et al.\ (\cite{mak2012}). Results of Doppler imaging reconstruction using data sets separated 
by just a couple of weeks  revealed the evolution of chemical spots already on 
such a short time scale (Briquet et al.\ \cite{Briquet2010}).
The Doppler maps computed by Briquet et al.\ (\cite{Briquet2010}) and Makaganiuk et al.\ (\cite{mak2012})
are roughly consistent, showing strongly overabundant Y and Sr patches at the rotation phases 0.75--1.00
and lower abundance patches around the phases 0.10--0.40. The Ti and Cr patches show a more complex 
surface distribution, although the distribution of the most overabundant patches roughly resembles 
those of Y and Sr. Furthermore, a significant variation in the latitudinal element distribution
and in the element abundance gradients is observed. Makaganiuk 
et al.\ (\cite{mak2012}) assumed the time of the first night of observation as the zero phase, which
corresponds to the rotation phase 0.42 in the work of Briquet et al.\ (\cite{Briquet2010}). 
As we mention above, the rotation phases presented in Table~\ref{tab:log_meas} for the star HD\,11753 
were calculated following the ephemeris used by  Briquet et al.\ (\cite{Briquet2010}).

\begin{figure}
\centering
\includegraphics[width=0.45\textwidth]{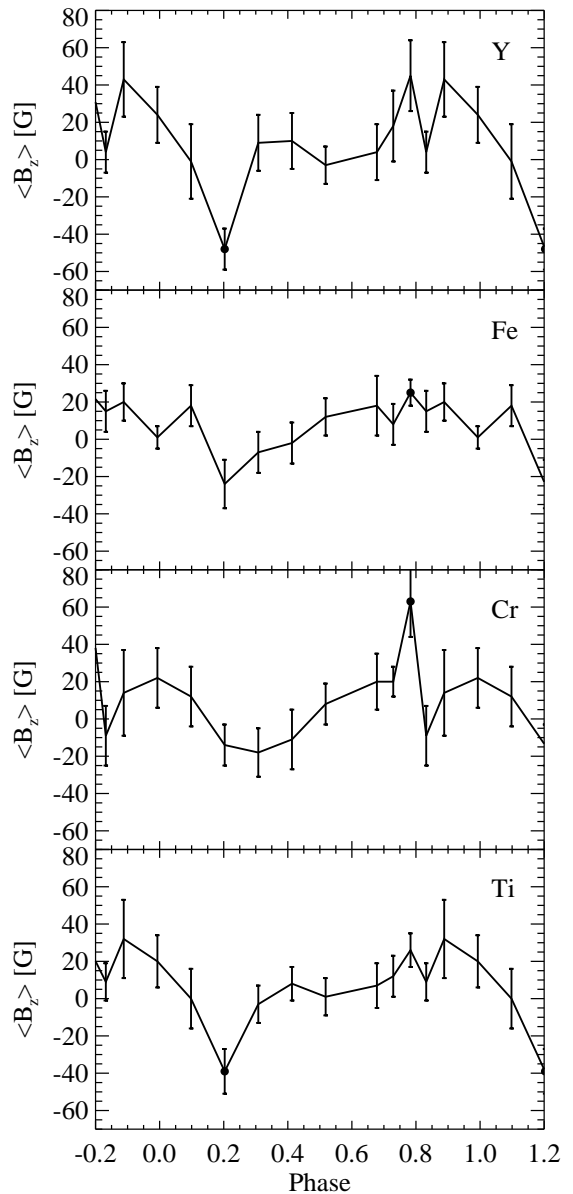}
\caption{
Measurements of the mean longitudinal magnetic field as a function of the rotation phase for HD\,11753.
The measurements were carried out separately for the elements Ti, Cr, Fe, and Y (from bottom to top). 
Filled circles indicate 3$\sigma$ measurements.
}
\label{fig:hd11753}
\end{figure}

The re-analysis of the HARPS spectra of this star by the moment technique using carefully selected line 
lists of the elements Ti, Cr, 
Fe, and Y shows remarkable results: The measurements on the Ti and Y lines reveal
the presence  of a weak negative magnetic field at the 3$\sigma$ significance level at rotation phase 0.203, while
the measurements on Cr and Fe show the presence of a weak positive magnetic field at the 3$\sigma$ significance level
at phase 0.783. Exactly at these rotation phases become the patches with the lowest surface element abundances as well as 
the patches with the highest surface element abundances best visible on the stellar surface.
Even more striking is the behaviour of the longitudinal magnetic field over the stellar surface
presented in Fig.~\ref{fig:hd11753}. The measurements on all four elements
indicate roughly the same kind of variability of the magnetic phase curve with predominantly negative
field polarity at the location of the low-abundance patches and positive polarity at the location of the 
high-abundance patches. The behaviour of the magnetic field over the rotation cycle looks
complex, indicating a non-sinusoidal character. It is also somewhat different 
for different elements and is very likely caused by differences in the surface abundance distribution 
for each element.
Such a result is not unexpected since the lines of 
different elements with different abundance distributions across the stellar 
surface sample the magnetic field in different manners. Also,
the presence of several element spots with different overabundances close to each other as presented 
in Fig.~8 in the work of Makaganiuk et al.\ (\cite{mak2012}) for the Ti distribution indicates a 
more complex underlying magnetic
field topology that cannot be represented by a simple dipole model.
The presence of tangled magnetic fields on the surface of this star is strengthened by our 3$\sigma$ detections 
of the mean quadratic magnetic field at two rotation phases, 0.413 and 0.993 (see Table~\ref{tab:log_meas2})
using Ti lines. As mentioned in the work of Makaganiuk et al.\ (\cite{mak2012}), this element shows 
the most complex surface element distribution. In addition, a 3$\sigma$ detection for positive crossover effect,
330$\pm$108\,km\,s$^{-1}$\,G, was achieved at the phase 0.832 for the sample of Ti lines, and for 
negative crossover effect,
$-$467$\pm$147\,km\,s$^{-1}$\,G, at the phase 0.678 for the sample of Fe lines.\\
%$v$\,sin\,$i$=15$\pm$8\,km\,s$^{-1}$?

{\it 41\,Eri:}
%\{bf THE ORDER OF SOME PARAGRAPHS HAS BEEN CHANGED }

With a visual magnitude of 3.55 this SB2 system is one of the brightest targets among our stellar sample,
with an excellent S/N achieved in the HARPS spectra. Four HARPS spectra were obtained for this star, but 
only three of them can be used for measurements because one observation was close to conjunction time, at
MJD\,55201.77, where the lines of both components overlap in the spectra. 
 
The atmospheric fundamental parameters for both components were studied by Dolk et al.\ (\cite{Dolk2003}),
who determined $T_{\rm eff}=12\,750$\,K for the primary and $T_{\rm eff}=12\,250$\,K for the secondary.
Our inspection of the spectra belonging to the primary and the secondary reveal that the line profiles
of several elements are clearly variable. Both components show typical HgMn peculiarities, but this is the first
time that spectrum variability was also discovered in the spectra of the secondary component. 

\begin{figure}
\centering
\includegraphics[angle=270,totalheight=0.35\textwidth]{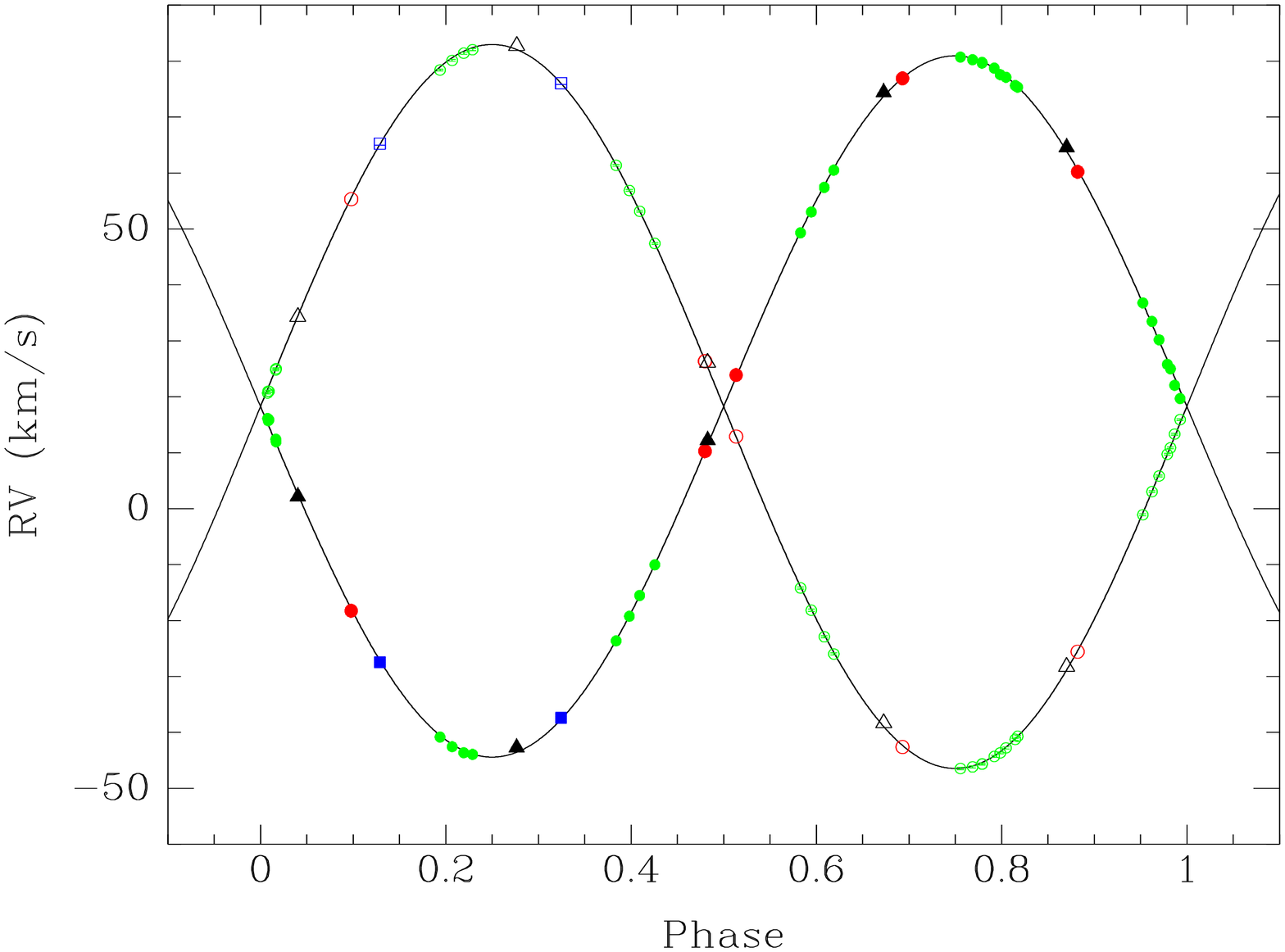}
\caption{
Spectroscopic orbit of the system 41\,Eri. Observations with FEROS are shown by red dots, 
with UVES by blue dots, with HARPS by black triangles, and with EBASIM by green symbols.
Open symbols indicate the radial velocity measurements in the spectra of the secondary component, which is 
the less massive star.}
\label{fig:hd27376orbit}
\end{figure}

\begin{table*}
\caption[]{
Radial velocity measurements carried out for the primary and secondary components
in the SB2 system 41\,Eri.
}
\label{tab:var_meas27376}
\centering
\begin{tabular}{lllr@{$\pm$}lr@{$\pm$}l}
\hline
\hline
\multicolumn{1}{c}{Instr.} &
\multicolumn{1}{c}{MJD} &
\multicolumn{1}{c}{Phase} &
\multicolumn{2}{c}{RV$_{\rm A}$} &
\multicolumn{2}{c}{RV$_{\rm B}$} \\
\hline
 FEROS  & 53663.2551 & 0.5134 &  23.88 & 0.13 &  12.89 & 0.22 \\
 FEROS  & 53664.1549 & 0.6930 &  76.92 & 0.15 & $-$42.61 & 0.25 \\
 FEROS  & 53665.1027 & 0.8821 &  60.21 & 0.12 & $-$25.57 & 0.23 \\
 FEROS  & 53666.1837 & 0.0979 & $-$18.26 & 0.14 &  55.32 & 0.23 \\
 FEROS  & 53668.0968 & 0.4797 &  10.29 & 0.13 &  26.37 & 0.22 \\
 FEROS  & 53668.0983 & 0.4800 &  10.27 & 0.13 &  26.35 & 0.22 \\
 UVES   & 53631.2672 & 0.1290 & $-$27.46 & 0.28 &  65.25 & 0.26 \\ % Prg. 076.D-0169(A)
 UVES   & 53632.2467 & 0.3245 & $-$37.40 & 0.25 &  76.06 & 0.23 \\
 EBASIM & 54398.1707 & 0.1938 & $-$40.84 & 0.24 &  78.41 & 0.18 \\
 EBASIM & 54398.2361 & 0.2069 & $-$42.56 & 0.24 &  80.13 & 0.16 \\
 EBASIM & 54398.2990 & 0.2194 & $-$43.64 & 0.26 &  81.42 & 0.17 \\
 EBASIM & 54398.3462 & 0.2289 & $-$43.94 & 0.29 &  82.05 & 0.18 \\
 EBASIM & 54399.1231 & 0.3839 & $-$23.63 & 0.26 &  61.37 & 0.17 \\
 EBASIM & 54399.1944 & 0.3982 & $-$19.24 & 0.24 &  56.87 & 0.16 \\
 EBASIM & 54399.2501 & 0.4093 & $-$15.52 & 0.24 &  53.16 & 0.16 \\
 EBASIM & 54399.3325 & 0.4257 & $-$10.03 & 0.26 &  47.40 & 0.16 \\
 EBASIM & 54400.1205 & 0.5830 &  49.32 & 0.23 & $-$14.19 & 0.17 \\
 EBASIM & 54400.1785 & 0.5946 &  53.04 & 0.25 & $-$18.16 & 0.19 \\
 EBASIM & 54400.2486 & 0.6086 &  57.43 & 0.24 & $-$22.90 & 0.20 \\
 EBASIM & 54400.3005 & 0.6189 &  60.51 & 0.25 & $-$25.98 & 0.18 \\
 EBASIM & 54401.1022 & 0.7790 &  79.77 & 0.26 & $-$45.64 & 0.17 \\
 EBASIM & 54401.2004 & 0.7985 &  77.59 & 0.27 & $-$43.67 & 0.17 \\
 EBASIM & 54426.2828 & 0.8047 &  77.12 & 0.28 & $-$42.77 & 0.18 \\
 EBASIM & 54426.3467 & 0.8174 &  75.34 & 0.24 & $-$40.70 & 0.18 \\
 EBASIM & 54427.0229 & 0.9524 &  36.78 & 0.25 &  $-$1.10 & 0.17 \\
 EBASIM & 54427.0727 & 0.9623 &  33.46 & 0.25 &   3.03 & 0.16 \\
 EBASIM & 54427.1111 & 0.9700 &  30.19 & 0.23 &   5.83 & 0.16 \\
 EBASIM & 54427.1729 & 0.9823 &  25.01 & 0.26 &  10.88 & 0.17 \\
 EBASIM & 54427.3072 & 0.0091 &  15.87 & 0.22 &  20.97 & 0.17 \\
 EBASIM & 54432.2353 & 0.9927 &  19.68 & 0.21 &  15.90 & 0.16 \\
 EBASIM & 54432.3133 & 0.0083 &  15.72 & 0.26 &  21.00 & 0.18 \\
 EBASIM & 54432.3558 & 0.0168 &  11.95 & 0.26 &  25.01 & 0.17 \\
 HARPS  & 55201.2686 & 0.4827 &  12.25 & 0.10 &  26.11 & 0.16 \\
 HARPS  & 55209.0727 & 0.0403 &   2.19 & 0.12 &  34.29 & 0.18 \\ 
 HARPS  & 55210.2560 & 0.2765 & $-$42.69 & 0.12 &  82.74 & 0.15 \\
 HARPS  & 55212.2414 & 0.6727 &  74.42 & 0.12 & $-$38.33 & 0.16 \\
 HARPS  & 55213.2305 & 0.8702 &  64.60 & 0.11 & $-$28.26 & 0.16 \\
% Note that this observation set has 3 pairs of spectra instead of 4.
%%FG Note I have removed the zeros of the last decimal in the HJD 
\hline
\end{tabular}
\tablefoot{
The HARPS observation at MJD\,55209.0727 from the ESO archive had just three sub-exposures and for this reason 
was not used in the magnetic field determination.
}
%\end{flushleft}
\end{table*}

\begin{table}
\caption{
Orbital and fundamental parameters for the SB2 system 41\,Eri.
%{\bf I changed km/s to km\,s$^{-1}$ in the table}
}
\label{tab:orbit}
\centering
\begin{tabular}{cr@{$\pm$}lc}
\hline
\hline
$P$ [d]                 & 5.01031535      &      0.00001217 \\
$T$ (MJD conj $\mathsc{i}$)         & 54407.220102      &      0.0012 \\
$V_{\rm o}$ [km\,s$^{-1}$]      & 18.27           &      0.07 \\
$K_{\rm A}$ [km\,s$^{-1}$]      & 62.68           &      0.17 \\
$K_{\rm B}$ [km\,s$^{-1}$]      & 64.70           &      0.15 \\
%$e$                     & \multicolumn{2}{c}{0}               & adopted \\
%$e$                     & 0.0001          &      0.0019      & as free param. \\
%%FG I moved the e to the text because it is not clear for me
%%FG if the rest of the parameters correspond to a fitting with eccentricity fixed
$a\,\sin\,i$ [$R_\odot$]     & 12.609          &      0.022 \\
$M_{\rm A}\,\sin^3\,i$ [$M_\odot$]  & 0.5450          &      0.0029 \\
$M_{\rm B}\,\sin^3\,i$ [$M_\odot$]  & 0.5279          &      0.0031 \\
$q$                          & 0.9688          &      0.0034 \\
\hline
$v\,\sin\,i_{\rm A}$ [km\,s$^{-1}$]  & 12.23           &      0.06\\
$v\,\sin\,i_{\rm B}$ [km\,s$^{-1}$]  & 11.78           &      0.07\\
$R_{\rm A}\,\sin\,i$ [$R_\odot$] & 1.199           &      0.010 \\
$R_{\rm B}\,\sin\,i$ [$R_\odot$] & 1.199           &      0.010 \\
%$R_{\rm A}/a	&	0.0960	&	0.0005 \\
%$R_{\rm B}/a	&	0.0925	&	0.0006 \\
%%FG these are R/a values calculated from vsini than can be added to this table if you wish.
\hline
\end{tabular}
\end{table}

The problem of analysing the component spectra in double-lined spectroscopic binaries is a tough one, but 
fortunately several techniques for spectral disentangling have been developed 
in the past few years. 
In this work, we have applied the procedure of decomposition described in detail by Gonz\'alez \& 
Levato (\cite{GonzalezLevato2006}).
The resulting spectra were used to study the spectral variability of the components and the
variation of radial velocities to improve the orbit of 41\,Eri. 
We used all high-resolution spectra at our disposal obtained with
%four different instruments, such as 
FEROS, UVES, and HARPS, and lower-resolution ($R=20\,000$) EBASIM spectra.
The ESO spectrographs UVES and FEROS are mounted on UT2 of the VLT at Paranal and
at the 2.2\,m telescope at La Silla, respectively. 
The EBASIM spectrograph is attached to the 2.1\,m Jorge Sahade telescope at the CASLEO in Argentina. 
All measured radial velocities used to fit the orbit 
of this system are presented in Table~\ref{tab:var_meas27376}, and the derived orbital parameters are listed in 
Table~\ref{tab:orbit}. 
In Fig.~\ref{fig:hd27376orbit}, we display the measured radial velocities together with the 
calculated orbital radial velocity curves. 
The orbit of the system 41\,Eri is indistinguishable from a circular orbit. Thus, the solution presented in Table~\ref{tab:orbit} corresponds
to an orbit fitting with fixed value $e=0$. 
If $e$ is fitted as a free parameter, we determine the value $e=0.0001\pm0.0019$.

The projected equatorial rotation velocities were measured using the mean
spectrum of each component obtained from the disentangling of HARPS spectra.
The parameter $v\,\sin\,i$ was calculated for several unblended \ion{Fe}{ii} and \ion{Cr}{ii} lines
through the measurement of the position of the first zero of the Fourier transform of
the line profiles. In these calculations, the appropriate linear limb darkening coefficient for the
corresponding stellar temperature and line wavelength was used.
Since the orbit is circular and the time-scale for circularisation is longer than that for
synchronisation, it can be assumed that the orbital motion and stellar rotation are
already synchronised. The projected stellar radii and the measured $v\,\sin\,i$-values for 
both components are presented in Table~\ref{tab:orbit}. 

The error assigned to the stellar rotation is the standard error of the mean
of the 25 lines measured and corresponds mainly to the random noise in the spectra.
A more realistic estimate of the uncertainties should include other sources of error.
We estimate that the influence of the instrumental profile is small, contributing with
about 0.05\,km\,s$^{-1}$. However, the limb darkening adopted for the rotational profile
is not well known, since its value may differ significantly from the continuum limb darkening.
For this reason, we assume that errors below 1\% for the rotational velocities 
are unrealistic and consequently adopted a 1\% error in the stellar radii as
a reasonable estimate.

%%FG Perhaps the rotation description is too long, but I needed to explain
%%FG why errors of 0.4% in radii seem to me too smaller.
%%FG If you prefer we can keep the smaller radii error and change the paragraph
%%FG to a warning saying that the errors might be somewhat larger.
In the disentangled spectra, the \ion{Hg}{ii} line at $\lambda$3984 and the \ion{Fe}{ii} lines appear to be stronger in the 
primary, while the Mn and  Ti lines are stronger in the less massive star, i.e.\ in the component B.
Since there is frequently confusion in the literature as to which component has to be considered as the primary
and which as the secondary, we add the information on the spectrum appearance directly in
both Tables~\ref{tab:log_meas} and \ref{tab:log_meas2}, which present the magnetic field measurements.

\begin{figure}
\centering
\includegraphics[angle=270,totalheight=0.35\textwidth]{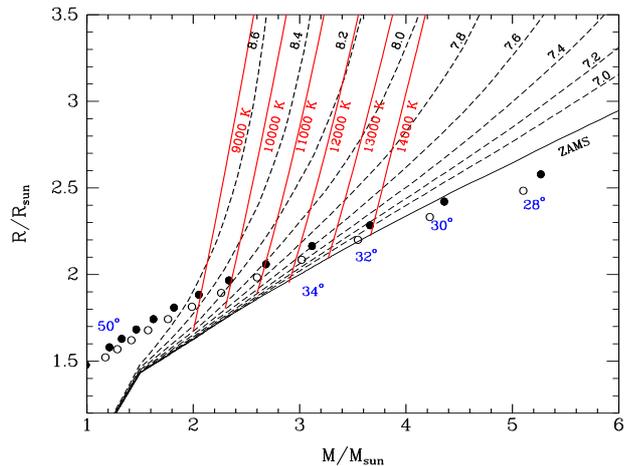}
\caption{
Position of the components in the system 41\,Eri in the mass--radius diagram for inclinations
in the range $28-50^\circ$ with a step of $2^\circ$. 
The red lines are isotherms interpolated in the stellar models.
With $T_{\rm eff}=12\,750$\,K for the primary and $T_{\rm eff}=12\,250$\,K for the secondary, we obtain 
an inclination of $33.7^\circ$.
%%Both components are located very close to the ZAMS. 
Open circles indicate the position of the secondary component.
%{\bf this figure has changed according to the new radii derived from new vsini}
}
\label{fig:hd27376hr}
\end{figure}

Further, we used a mass--radius (M--R) diagram to estimate the inclination of the system
and the absolute stellar parameters.
In Fig.~\ref{fig:hd27376hr}, we show the position of both components of 41\,Eri in the
M--R diagram for various possible values of orbital inclination, together with several isochrones and isotherms
from stellar models by Schaller et al. (\cite{Schaller1992}).
%Several isochrones and isotherms for theoretical stellar models have been plotted. }
Assuming $T_{\rm eff}=12\,750$\,K for the primary and $T_{\rm eff}=12\,250$\,K for the secondary 
(see e.g.\ Dolk et al.\ \cite{Dolk2003}), we  obtain an inclination of $33.7^\circ$. 
%The apparent difference in age in both systems is due
%to the fact that $v\,\sin\,i$-value used in the M-R diagram is assumed to be the same for both
%components.  (Dolk et al.\ \cite{Dolk2003}). 
The age of the system is estimated to be around 25--30 Myr, representing less than
10\% of its main sequence life time.
%? The small difference in radius expected according
%? to the isochrones might be compatible with errors in $v\,\sin\,i$-values. 

%{\bf Federico, how the 0.1 km/s  accuracy can cause a big difference in the radii?}
%%FG the  relative error of radii and vsini are similar.
%%FG The formal error of vsini are now very small. Probably other error sources
%%FG related with the instrumental line profile and limb darkening would also
%%FG contribute but I have not estimate them yet.
%We note that the very precise 
%measurements of $v\,\sin\,i$-values in both components is difficult due to the presence of the inhomogeneous 
%element distribution and very narrow cores in some phases. 

We note that the system 41\,Eri actually seems to be a triple one:
Hubrig et al.\ (\cite{Hubrig2001}) 
studied this system with diffraction-limited near-infrared ADONIS
observations on La~Silla and found a companion of K = 9.9 at a distance of 5\farcs32 and a position 
angle of 162.5$^\circ$.
This finding was later confirmed by observations of Sch\"oller et al.\ (\cite{Schoeller2010}) with 
NAOS-CONICA at the VLT. 

\begin{figure*}
\centering
\includegraphics[angle=270,width=0.49\textwidth]{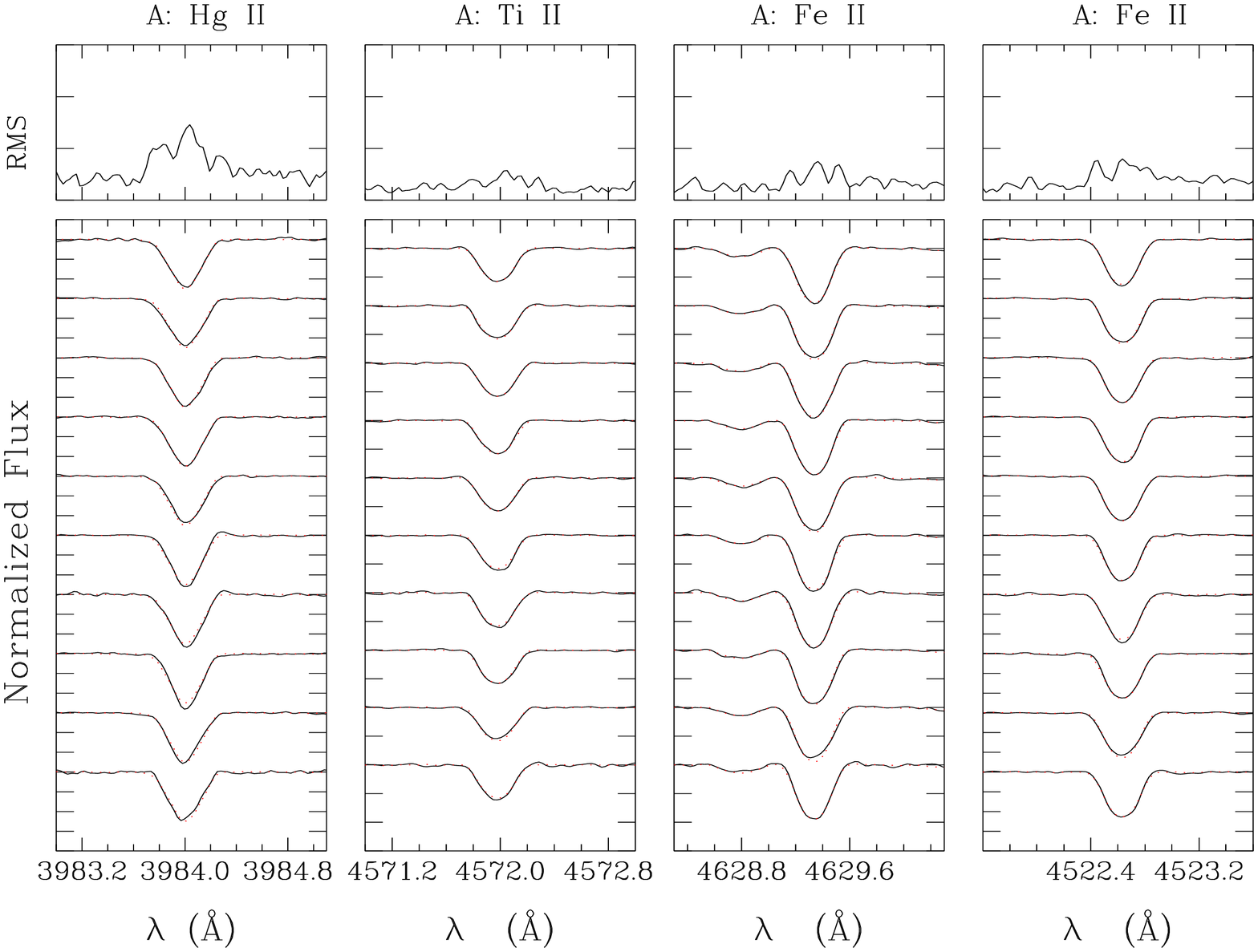}
\includegraphics[angle=270,width=0.49\textwidth]{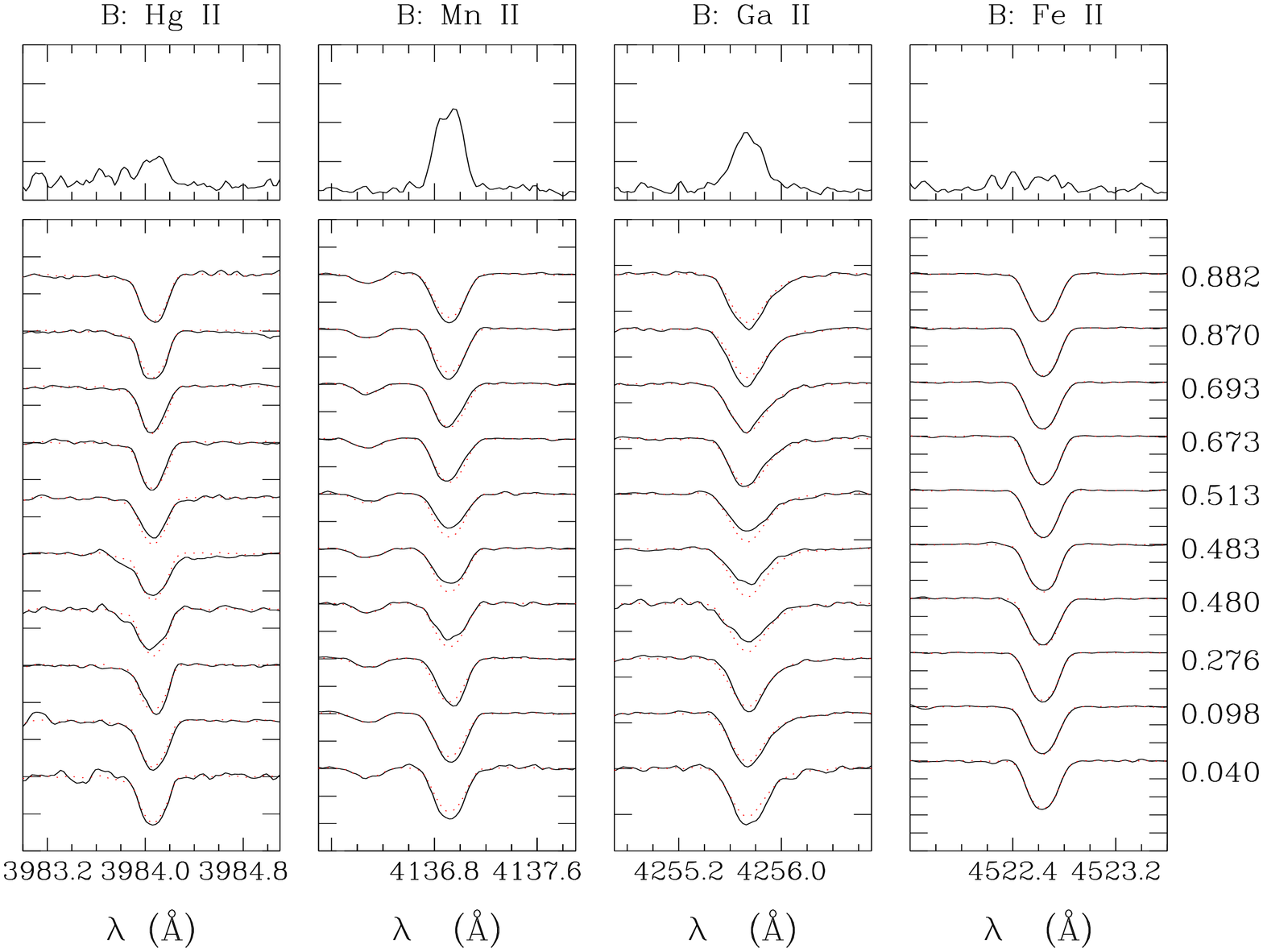}
\caption{
%{\bf figure has been remade for a better visualisation}
Line-profile variations and the RMS values in the spectra of both components of 41\,Eri over the rotation period.
The dotted line corresponds to the mean line profiles. The scales for the normalised flux 
have ticks of 0.1 in the lower panels and 0.01 in the upper panels, respectively.
}
\label{fig:hd27376_var}
\end{figure*}

The variability of the spectral lines was studied in the high-resolution disentangled spectra
for each orbital phase.  
A distinct variability is observed for Hg, Ti, and Fe lines in the primary and for Hg, Mn, Ga, and Fe lines 
in the secondary. 
The profile variations of several lines over the rotation cycle are presented in Fig.~\ref{fig:hd27376_var}.
%{\bf Federico, the x-axes have to be improved in these plots.}
In the upper panel, we present the root mean square (RMS) values of the
residuals. The strongest RMS peaks correspond to the Hg line in the spectra of the primary and to Mn and Ga lines in the spectra 
of the secondary component.

Since in both stars the line profiles 
belonging to several
elements show significant variability, the disentangling process requires
a dense high-resolution spectral time series. We note that even with the rather dense spectral time series
obtained, 
the results are less reliable in spectral regions where both components exhibit variable line profiles.
The analysis of the Mn line-profile variations in the secondary indicates that
Mn is concentrated in a spot located on the stellar hemisphere opposite to the hemisphere facing the companion star.
%in the face that is hidden to the companion (external face). 
Around the orbital phase 0.5, i.e.\  the phase when the stellar surface facing the primary
becomes better visible, the Mn lines appear significantly weaker. The Ga lines and the \ion{Hg}{ii} line
at $\lambda$3984 show qualitatively a very similar behaviour. 
%In any case this system appears to be an excellent target for future Doppler Imaging studies 
%to probe the correlations between the binary properties, magnetic field structure, and 
%abundance inhomogeneities.    

\begin{figure}
\centering
\includegraphics[width=0.45\textwidth]{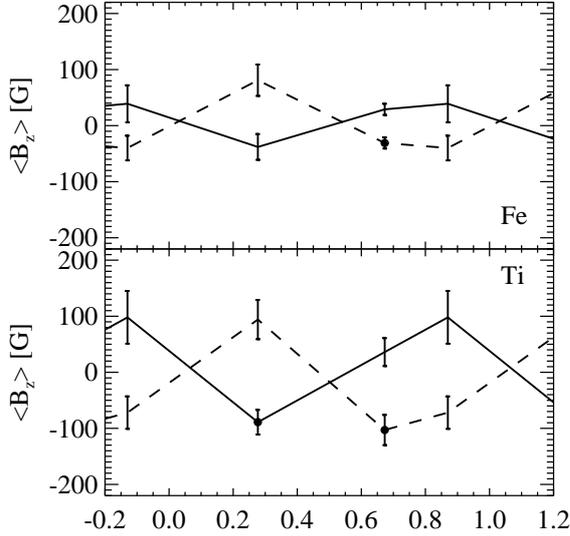}
\caption{
Measurements of the mean longitudinal magnetic field as a function of the rotation phase for 41\,Eri.
The measurements were carried out separately for the elements Ti (bottom) and Fe (top). 
The solid line denotes the primary component, while the dashed line denotes the secondary component.
Filled circles indicate 3$\sigma$ measurements.
}
\label{fig:hd27376}
\end{figure}

The distribution of the available measurements of the mean longitudinal magnetic field over the stellar 
surface in both components is presented  in Fig.~\ref{fig:hd27376}. As mentioned above, 
due to the overlap of
spectral lines of components in one of the HARPS spectra, reliable results of the measurements 
can only be achieved at three rotation phases.
$3\sigma$ detections were achieved using the sample of Ti lines in the spectra of the primary 
at the rotation phase 
0.276 as well as the samples of Ti and Fe lines in the spectra of the secondary at the phase 0.673. 
An interesting fact discovered in these measurements
is that the stellar surfaces with low-abundance element spots that face the companion star show negative 
magnetic field polarity, while 
for the opposite hemisphere covered by high abundance  
element spots, the magnetic field is positive.
% around the rotation phase of best spot visibility.  

A mean quadratic
magnetic field at the 3$\sigma$ level, 4520$\pm$758\,G, was detected in the primary at the phase 0.870  
using the sample of Ti lines.
As for the secondary, a quadratic magnetic field of 2332$\pm$488\,G and 2106$\pm$634\,G was detected at the phases 
0.673 and 0.870, respectively, using the sample of Fe lines. 
Furthermore, we detect a quadratic
magnetic field at a 3$\sigma$ level, 3885$\pm$1115\,G at the phase 0.673, and a quadratic magnetic field 
4484$\pm$1201\,G at the phase 0.870 using Ti lines.
No crossover effect at a 3$\sigma$ level was detected for this system.

{\it 66\,Eri:}
%{\bf see comment in tex}
%%FG I was wrong when I told you that the HgMn star was the less massive.
%%FG Star A, the most massive and larger is the HgMn. This fact does
%%FG not change  anything in the paper text and
%%FG does not mean that Makaganiuk does not have internal identification problems
%%FG (their K_A K_B have to be swapped).

\begin{table*}
\caption[]{
Measurements of radial velocities for the primary and secondary component
of the SB2 system 66\,Eri. 
}
\label{tab:var_meas32964}
\centering
\begin{tabular}{lccr@{$\pm$}lr@{$\pm$}l}
\hline
\hline
%HJD-2400000  pha_con pha_per  rva     erra   rvb    errb
\multicolumn{1}{c}{HJD} &
\multicolumn{1}{c}{Phase} &
\multicolumn{1}{c}{Phase} &
\multicolumn{2}{c}{RV$_{\rm A}$} &
\multicolumn{2}{c}{RV$_{\rm B}$} \\
\multicolumn{1}{c}{$-$2,400,000} &
\multicolumn{1}{c}{Conjunction} &
\multicolumn{1}{c}{Period} &
\multicolumn{2}{c}{[km\,s$^{-1}$]} &
\multicolumn{2}{c}{[km\,s$^{-1}$]} \\
\hline
%CORALIE
\multicolumn{7}{c}{CORALIE} \\
\hline
 55212.5706 & 0.69826 & 0.97411 & 127.32 & 0.31 & $-$65.82 & 0.23  \\
 55212.7135 & 0.72413 & 0.99998 & 135.30 & 0.31 & $-$73.73 & 0.24  \\
 55213.5546 & 0.87644 & 0.15228 & 111.55 & 0.34 & $-$49.47 & 0.23  \\
 55213.7138 & 0.90526 & 0.18111 &  96.41 & 0.32 & $-$34.28 & 0.22  \\
 55214.5572 & 0.05798 & 0.33383 &   6.41 & 0.31 &  58.08 & 0.22  \\
 55214.7184 & 0.08717 & 0.36302 &  $-$8.67 & 0.30 &  73.70 & 0.22  \\
 55215.6575 & 0.25722 & 0.53306 & $-$59.42 & 0.34 & 125.70 & 0.22  \\
 55216.5658 & 0.42169 & 0.69753 & $-$32.08 & 0.34 &  97.97 & 0.22  \\
 55216.7108 & 0.44794 & 0.72379 & $-$20.15 & 0.33 &  86.00 & 0.22  \\
 55217.5666 & 0.60291 & 0.87875 &  75.44 & 0.35 & $-$12.78 & 0.24  \\
 55217.6999 & 0.62704 & 0.90289 &  91.05 & 0.33 & $-$28.66 & 0.24  \\
 55220.5940 & 0.15109 & 0.42694 & $-$36.04 & 0.35 & 101.58 & 0.25  \\
 55220.7149 & 0.17298 & 0.44883 & $-$43.05 & 0.35 & 108.94 & 0.27  \\
 55224.6002 & 0.87651 & 0.15236 & 111.45 & 0.34 & $-$49.39 & 0.26  \\
 55224.6936 & 0.89342 & 0.16927 & 102.68 & 0.38 & $-$40.71 & 0.29  \\
 55226.5722 & 0.23359 & 0.50944 & $-$56.75 & 0.33 & 123.01 & 0.23  \\
 55230.5862 & 0.96040 & 0.23627 &  64.04 & 0.31 &  $-$0.85 & 0.23   \\
 55230.6666 & 0.97496 & 0.25083 &  55.32 & 0.31 &   8.38 & 0.23  \\
\hline
%HARPS
\multicolumn{7}{c}{HARPS} \\
\hline
 55202.7262 & 0.91569 & 0.19154 &  91.50 & 0.32 & $-$27.64 & 0.18  \\
 55203.7457 & 0.10030 & 0.37614 & $-$14.14 & 0.29 &  80.64 & 0.18  \\
 55204.7384 & 0.28005 & 0.55590 & $-$59.66 & 0.33 & 127.41 & 0.19  \\
 55205.7570 & 0.46449 & 0.74034 & $-$11.38 & 0.33 &  78.25 & 0.17  \\
 55206.7375 & 0.64203 & 0.91788 & 100.79 & 0.33 & $-$37.08 & 0.18  \\
 55207.7363 & 0.82289 & 0.09874 & 133.31 & 0.33 & $-$70.24 & 0.18  \\
 55209.7426 & 0.18618 & 0.46203 & $-$46.10 & 0.31 & 113.46 & 0.18  \\
 55210.7476 & 0.36816 & 0.64401 & $-$49.21 & 0.32 & 116.75 & 0.18  \\
 55211.7523 & 0.55009 & 0.82593 &  40.88 & 0.33 &  24.85 & 0.17  \\
 55212.7290 & 0.72694 & 0.00279 & 136.63 & 0.30 & $-$73.83 & 0.17  \\
\hline
%UVES
\multicolumn{7}{c}{UVES} \\
\hline
 53632.8917 & 0.65913 & 0.93498 & 109.56 & 1.07 & $-$47.27 & 1.29  \\
 53662.7238 & 0.06096 & 0.33681 &   4.35 & 1.22 &  59.01 & 1.41  \\
 53662.7254 & 0.06125 & 0.33710 &   4.25 & 1.23 &  59.21 & 1.41  \\
\hline
%FEROS
\multicolumn{7}{c}{FEROS} \\
\hline
 53663.8064 & 0.25699 & 0.53284 & $-$59.04 & 0.92 & 126.52 & 1.12  \\
 53664.6783 & 0.41487 & 0.69072 & $-$34.66 & 1.04 & 101.19 & 1.15  \\
 53665.6415 & 0.58928 & 0.86513 &  66.72 & 0.91 &  $-$3.18 & 1.04  \\
 53666.7037 & 0.78161 & 0.05746 & 139.46 & 0.93 & $-$77.86 & 1.13  \\
\hline
\end{tabular}
\end{table*}

\begin{table*}
\caption[]{
Measurements of radial velocities for the primary and secondary component
of the SB2 system 66\,Eri, taken from the literature.
}
\label{tab:var_meas32964_2}
\centering
\begin{tabular}{lccr@{$\pm$}lr@{$\pm$}l}
\hline
\hline
%HJD-2400000  pha_con pha_per  rva     erra   rvb    errb
\multicolumn{1}{c}{HJD} &
\multicolumn{1}{c}{Phase} &
\multicolumn{1}{c}{Phase} &
\multicolumn{2}{c}{RV$_{\rm A}$} &
\multicolumn{2}{c}{RV$_{\rm B}$} \\
\multicolumn{1}{c}{$-$2,400,000} &
\multicolumn{1}{c}{Conjunction} &
\multicolumn{1}{c}{Period} &
\multicolumn{2}{c}{[km\,s$^{-1}$]} &
\multicolumn{2}{c}{[km\,s$^{-1}$]} \\
\hline
%Young (1976) PASP 88,275 (errors modified)
\multicolumn{7}{c}{Young (\cite{Young1976})} \\
\hline
 41379.6890 & 0.92001 & 0.19586 &  84.60 & 2.50 & $-$28.30 & 3.00  \\
 41380.7080 & 0.10452 & 0.38037 & $-$15.10 & 2.50 &  77.60 & 3.00  \\
 41381.6830 & 0.28107 & 0.55692 & $-$62.30 & 2.50 & 125.50 & 3.00  \\
 41382.6700 & 0.45979 & 0.73564 & $-$17.80 & 2.50 &  81.80 & 3.00  \\
 41383.6870 & 0.64394 & 0.91979 & 101.80 & 2.50 & $-$40.10 & 3.00  \\
 42332.0070 & 0.36025 & 0.63610 & $-$52.80 & 2.50 & 122.80 & 3.00  \\
 42333.9840 & 0.71823 & 0.99408 & 134.60 & 2.50 & $-$69.90 & 3.00  \\
\hline
%Yushchenko et al (1999) Astron. lett. 25, 453
\multicolumn{7}{c}{Yushchenko et al.\ (\cite{Yushchenko1999})} \\
\hline
 50057.3750 & 0.22523 & 0.50108 & $-$53.00 & 1.50 & 123.10 & 1.50  \\
 50060.4060 & 0.77407 & 0.04992 & 142.90 & 1.50 & $-$74.80 & 1.50  \\
\hline
%Yushchenko et al (2001) IBVS 5213 (errors modified)
\multicolumn{7}{c}{Yushchenko et al.\ (\cite{Yushchenko2001})} \\
\hline
 50689.9890 & 0.77532 & 0.05117 & 139.70 & 1.50 & $-$79.50 & 1.50  \\
 50690.9820 & 0.95513 & 0.23098 &  67.00 & 1.50 &  $-$4.60 & 1.50  \\
 50691.9900 & 0.13765 & 0.41350 & $-$30.50 & 1.50 &  96.30 & 1.50  \\
 50692.9860 & 0.31800 & 0.59385 & $-$59.40 & 1.50 & 124.60 & 1.50  \\
 50693.9770 & 0.49745 & 0.77329 &   5.30 & 1.50 &  57.70 & 1.50  \\
\hline
%Catanzaro & Leto (2004) A&A 416,661
\multicolumn{7}{c}{Catanzaro \& Leto (\cite{CatanzaroLeto2004})} \\
\hline
 52189.5966 & 0.31560 & 0.59145 & $-$64.50 & 3.80 & 121.80 & 2.10  \\
 52202.5951 & 0.66929 & 0.94514 & 113.80 & 1.40 & $-$52.80 & 2.00  \\
 52204.5894 & 0.03041 & 0.30626 &  31.70 & 1.00 &  31.70 &  1.00  \\
 52205.5665 & 0.20734 & 0.48318 & $-$52.00 & 1.80 & 115.40 & 2.20  \\
 52220.5585 & 0.92200 & 0.19785 &  85.40 & 1.80 & $-$25.30 & 1.40  \\
 52221.5053 & 0.09344 & 0.36929 & $-$11.30 & 1.30 &  78.00 & 1.50  \\
 52235.5245 & 0.63196 & 0.90781 &  95.20 & 2.60 & $-$34.80 & 0.80  \\
 52250.4455 & 0.33377 & 0.60961 & $-$57.40 & 1.10 & 123.10 & 2.20  \\
 52566.5808 & 0.57772 & 0.85357 &  59.60 & 0.90 &   6.20 & 1.90  \\
 52567.5663 & 0.75617 & 0.03202 & 141.80 & 1.20 & $-$74.50 & 1.80  \\
\hline
\end{tabular}
\end{table*}

Ten HARPS polarimetric spectra were obtained over almost two orbital cycles of this SB2 system.
Among them, one spectrum was taken close to the time of conjunction $\mathsc{ii}$, at
MJD\,55211.25, where the lines of both components overlap in the spectra. 
The most recent determination of fundamental and orbital parameters of 66\,Eri was carried out by Makaganiuk et 
al.\ (\cite{Makaganiuk2011a}). In their work, however, there is a mismatch in the presentation of 
the orbital and fundamental parameters of the system, i.e.\ some parameters are ascribed to component A, 
while in actuality
they belong to component B.  We re-analysed 66\,Eri using all spectroscopic material
available to us: ten HARPS, four FEROS, and three UVES spectra.
Furthermore, 18 spectra at a resolution of $\sim60\,000$ were obtained with the CORALIE echelle spectrograph,
attached to the 1.2\,m Leonard Euler telescope 
on La~Silla in Chile. Additional radial velocity measurements were compiled from the literature.
All these measurements for both the primary and the secondary component are presented in 
Tables~\ref{tab:var_meas32964} and \ref{tab:var_meas32964_2}. 
Similar to the study of 41\,Eri, these radial velocity measurements were obtained as a 
result of spectral disentangling.
As mentioned above, our phase zero for 66\,Eri refers to the time of conjunction $\mathsc{i}$, while the
phase zero in the work of Makaganiuk et al.\ (\cite{Makaganiuk2011a}) refers to time of periastron.
The time of conjunction is more useful because it refers to the position with respect to the 
companion, while the time of periastron is generally not 
well defined for binary systems with non-zero eccentricities. 
%while the time of conjunction refers to the position with respect to the companion. 
Furthermore, our MJD values are slightly different (by about 0.0012) 
from those presented by Makaganiuk et al.\ (\cite{Makaganiuk2011b}) because they are heliocentric and 
correspond to the middle of exposures.% and not to the beginning  of the exposures.

\begin{figure*}
\centering
\includegraphics[angle=270,width=0.48\textwidth,totalheight=0.5\textwidth]{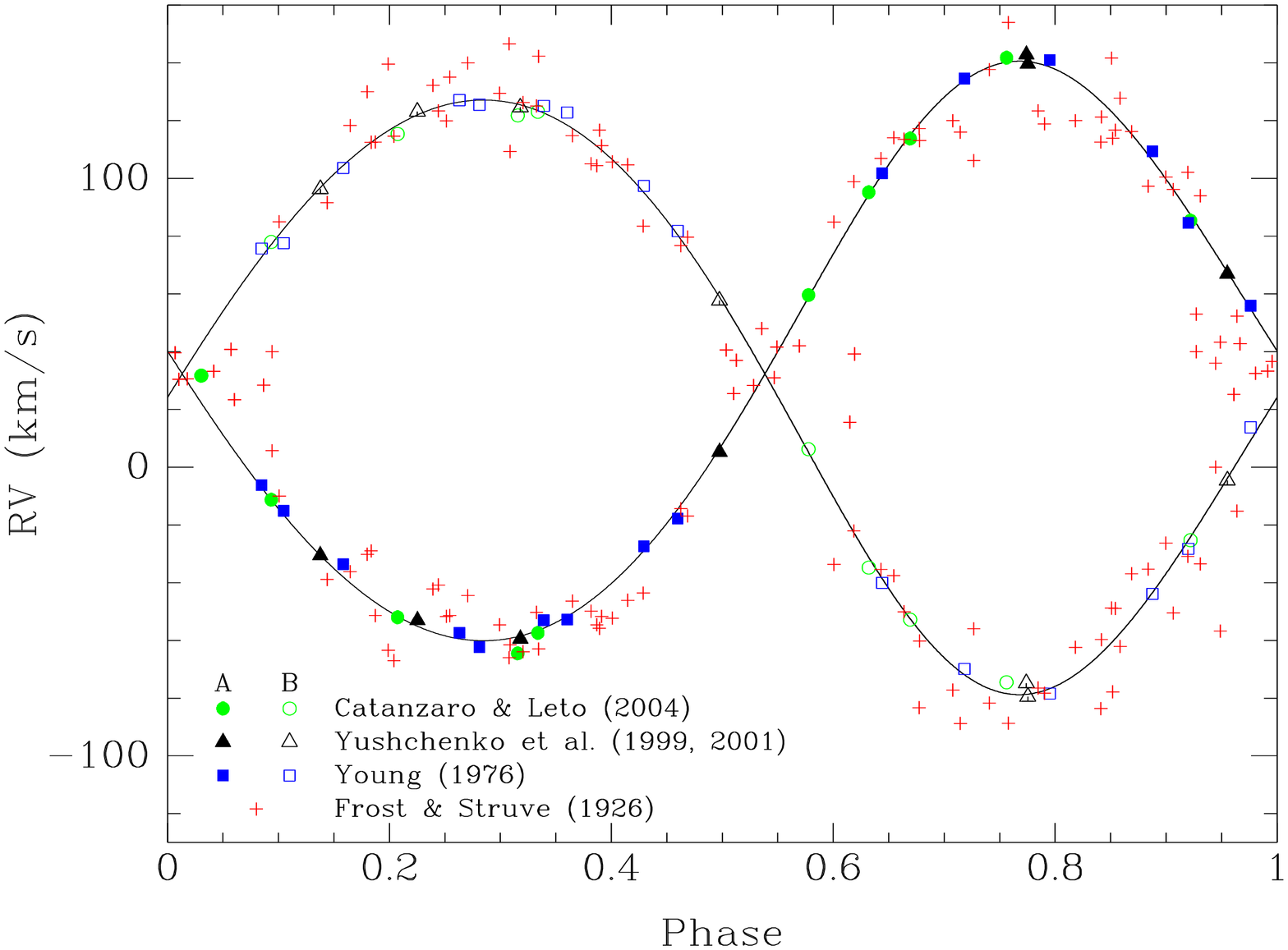}
\includegraphics[angle=270,width=0.48\textwidth,totalheight=0.5\textwidth]{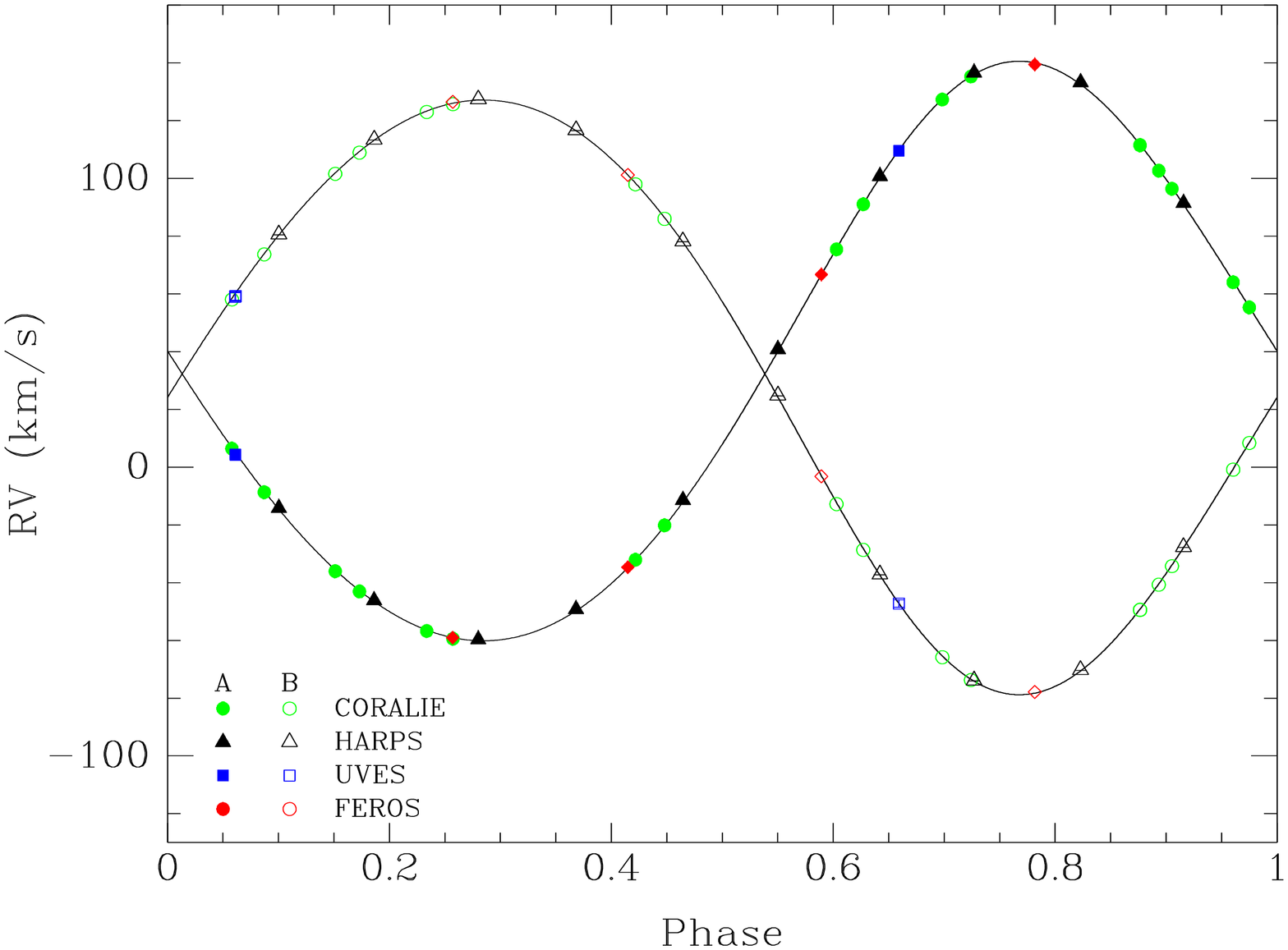}
\caption{
Spectroscopic orbit of the system 66\,Eri obtained using old data from the literature (left side) and new data 
consisting of ten HARPS, four FEROS, three UVES, and 18 CORALIE spectra (right side).
Open symbols indicate the measurements of the secondary component, which is the less massive star.
}
\label{fig:hd32964orbit}
\end{figure*}

\begin{table}
\caption{
Orbital and fundamental parameters for the SB2 system 66\,Eri.
%{\bf I changed $K_1$ $K_2$ to $K_A$ $K_B$, and km/s to km\,s$^{-1}$ in the table}
}
\label{tab:orbit32964}
\centering
\begin{tabular}{cr@{$\pm$}lc}
\hline
\hline
%old
%P [d]           5.5226057       +-      0.0000080
%T(MJD conjI)    53661.88467     +-      0.0032
%T(MJD periast)  53660.36107     +-      0.0034
%Vo (km/s)      32.64           +-      0.08
%K_A (km/s)     100.44           +-      0.21
%K_B (km/s)     102.95           +-      0.12
%e               0.0836          +-      0.0011         
%asini (Rsun)    22.114          +-      0.026
%M_A sin3i(Msun) 2.411           +-      0.008 
%M_B sin3i(Msun) 2.353           +-      0.010 
%q               0.9756          +-      0.0023
%
$P$ [d]                 & 5.5225973 & 0.000002 \\
$T$ (conj)         & 2455208.714 & 0.002 \\
$T$ (per)          & 2455207.191 & 0.010 \\
$V_{\rm o}$ [km\,s$^{-1}$]      & 32.30 & 0.07 \\
$K_A$ [km\,s$^{-1}$]      & 100.36 & 0.18 \\
$K_B$ [km\,s$^{-1}$]      & 102.97 & 0.12 \\
$\omega$          & 5.960 & 0.011 \\
$e$                     & 0.0833 & 0.0010 \\
$a\,\sin\,i$ [$R_\odot$]     & 22.109 &0.025 \\
$M\,\sin^3\,i$ [$M_\odot$]  & 4.761 & 0.016 \\
$M_{\rm A}\,\sin^3\,i$ [$M_\odot$]  &2.411 &  0.007 \\
$M_{\rm B}\,\sin^3\,i$ [$M_\odot$]  &2.350 & 0.009 \\
$q$                          & 0.9747 & 0.0022 \\ \hline
$v\,\sin\,i_{\rm A}$ [km\,s$^{-1}$]  & 17.68           &      0.04\\
$v\,\sin\,i_{\rm B}$ [km\,s$^{-1}$]  & 17.00           &      0.05\\
$R_{\rm A}\,\sin\,i$ [$R_\odot$] &       1.85   &     0.02       \\
$R_{\rm B}\,\sin\,i$ [$R_\odot$] &       1.78   &     0.02      \\
\hline
\end{tabular}
\end{table}

In addition, we used the 83 old radial velocity measurements from Frost \& Struve (\cite{Frost1924}).
The orbital parameters are shown in Table~\ref{tab:orbit32964}. 
In Fig.~\ref{fig:hd32964orbit},
we present our calculated radial velocity curves along with the old and new radial velocity data.

%%FG I do not see much difference with Makaganiuk in the parameters
%Note that some differences do exist 
%between our determinations and those of Makaganiuk et al.\ (\cite{Makaganiuk2011a}).
%{\bf Federico, which luminosity ratio have you assumed, say around the Mg II 4481 line and in the blue and red? 
%Which vsini values you assumed for calculation of fractional radii in 66 Eri? Why the vsini for the secondary
%in the work of Mak. is lower than the vsini for the primary?}
%%FG vsini have been remeasured  with FOurier Transform
Due to a rather large eccentricity of the system, $e=0.0844$, the rotation period of 66\,Eri 
is not expected to be the same as the orbital period. 
The period of corotation at periastron is 4.657\,d, while the pseudo-synchronous period
as defined by Hut (\cite{hut81}), is 5.3015\,d. 
The latter is expected to correspond to the rotation period.
Makaganiuk et al.\ (\cite{Makaganiuk2011a}) assumed  
the period of the orbital motion as the period of spectral 
variability, while our rotation phases presented in Table~\ref{tab:log_meas}
adopt the pseudo-synchronisation period.
%Thus we assume that the orbital period in this system is the same as the rotation period of 
%the components.

\begin{figure}
\centering
\includegraphics[angle=270,totalheight=0.35\textwidth]{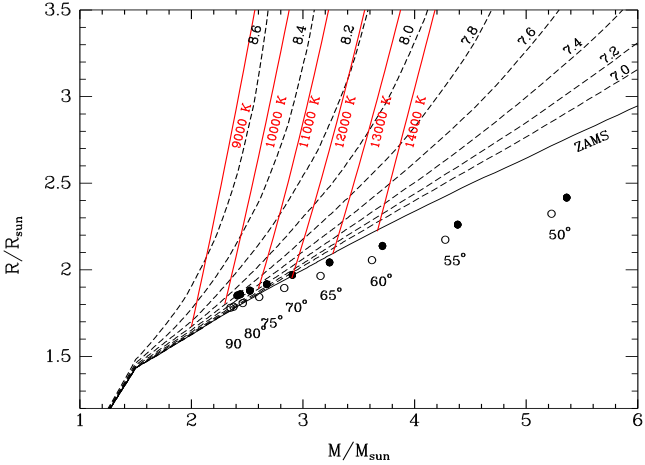}
\caption{Position of the components of 66\,Eri in the M--R diagram for inclinations
in the range $50-90^\circ$ with a step of $2^\circ$. 
%The red lines are isotherms interpolated in the stellar models.
With $T_{\rm eff}=11\,077$\,K for the primary and $T_{\rm eff}=10\,914$\,K for the secondary, we obtain 
an inclination of $73.7^\circ$.
Both components are located very close to the ZAMS. Open circles indicate the position of the 
secondary component.
%{\bf this figure has changed according to the new radii derived from new vsini}
}
\label{fig:hd32964hr}
\end{figure}

Similar to the procedure applied for 41\,Eri, we used the M--R diagram to estimate the 
inclination of the system.
Assuming $T_{\rm eff}=11\,077$\,K for the primary and $T_{\rm eff}=10\,914$\,K for the secondary 
(see e.g.\ Makaganiuk et al.\ \cite{Makaganiuk2011a}),
the system appears at an inclination of $73.7^\circ$. 
Figure \ref{fig:hd32964hr} shows the position of the stellar components in the
 M--R diagram, where both companions are located very close to the zero age main sequence (ZAMS). 
%The small difference in radius expected according
%to the isochrones might be compatible with errors in $v\,\sin\,i$-values. 

The presence of an additional companion in this system was reported from diffraction-limited near-infrared
observations.
Hubrig et al.\ (\cite{Hubrig2001}) 
studied this system with the diffraction-limited near-infrared ADONIS system
on the ESO 3.6\,m telescope and found a companion of K = 9.4 at a distance of 1\farcs613 and a 
position angle of 232.6$^\circ$.
The presence of a companion was later confirmed by observations of Sch\"oller et al.\ (\cite{Schoeller2010})
with NAOS-CONICA at the VLT. 

\begin{figure}
\centering
\includegraphics[width=0.45\textwidth]{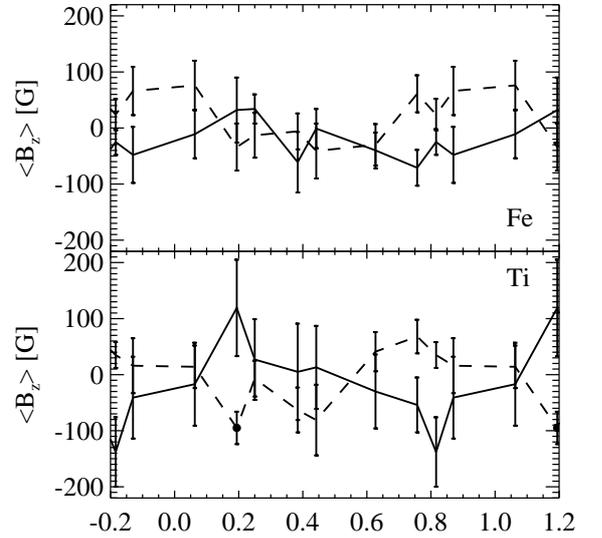}
\caption{
Measurements of the mean longitudinal magnetic field as a function of the rotation phase for 66\,Eri.
The measurements were carried out separately for the elements Ti and Fe. 
The solid line denotes the primary component, while the dashed line denotes the secondary component.
Filled circles indicate 3$\sigma$ measurements.
}
\label{fig:hd32864}
\end{figure}

The quality of the polarimetric HARPS spectra is not as good as the spectra of HD\,11753 and 41\,Eri,
with S/N mostly between 200 and 300. This could explain the larger inaccuracies in the magnetic 
field determinations and the fact that in our study only one detection at a 3$\sigma$ significance level was 
achieved at the rotation phase 0.186 using Ti lines. 
The distribution of the mean longitudinal magnetic field values measured in both components 
over the rotation cycle is presented in Fig.~\ref{fig:hd32864}. A comparison of the field distribution with 
the distribution of elements over the stellar surface presented in Fig.~7 in the work of Makaganiuk et 
al.\ (\cite{Makaganiuk2011a}) confirms the pattern already discovered in 
HD\,11753 and 41\,Eri: A negative mean longitudinal magnetic field is measured 
at the location of the lower abundance patches, i.e.\ on the stellar surface facing the companion, and 
the positive mean longitudinal magnetic field roughly corresponds to the 
location of the high-abundance patches located on the opposite hemisphere.
A positive 3$\sigma$ crossover, 1116$\pm$374\,km\,s$^{-1}$\,G, is observed in the sample of Fe lines in the 
primary at the phase 0.384.
Using Fe lines in the secondary, we obtain a positive crossover 1795$\pm$396\,km\,s$^{-1}$\,G, at the phase 0.442. 
Furthermore, negative crossover, $-$1382$\pm$450\,km\,s$^{-1}$\,G, is observed in Fe lines in the primary
at the phase 0.816. Interestingly, no significant crossover effect was found either in the primary or 
in the secondary using the sample of Ti lines.
A mean quadratic magnetic field at the 3$\sigma$ significance level was discovered in both components at several 
rotation phases using samples of Ti and Fe lines. 
%in phases 0.207 and 0.918 for the A (SYMMETRIC in the orbital curve!),
%and in phases 0.054, 0.133, 0.274, 0.564, 0.781, 0.918 in the B component - almost everywhere 

{\it HD\,33904:}

Until now, four polarimetric observations have been obtained with HARPS for this HgMn star, but only one
observation was publically available in the ESO archive at the time of our visitor stay at the ESO headquarters,
which was aimed at the reduction of the HARPS spectra with the HARPS pipeline machine.
The spectral variability of several elements was studied by Kochukhov et al.\ (\cite{Koch2011}), who also
used the LSD technique with the mask covering 526 lines to determine the longitudinal magnetic field and
concluded that the upper limit for the field in this star is only about 3\,G.
The available radial velocity data are not sufficient to decide whether this star is a member of a SB system,
but recent adaptive optics observations by Sch\"oller et al.\ (\cite{Schoeller2010}) presented the first 
direct detection of a close companion candidate at a separation of 0\farcs352 and a position angle of 
250.9$^{\circ}$.
Measurements using the sample of
Ti lines reveal that the mean quadratic magnetic field is definitely present on the surface of this star, while
the mean longitudinal magnetic field measured on the same lines accounts for a 2.9$\sigma$ detection.

{\it AR\,Aur:}

\begin{figure}
\centering
\includegraphics[angle=0,totalheight=0.65\textwidth]{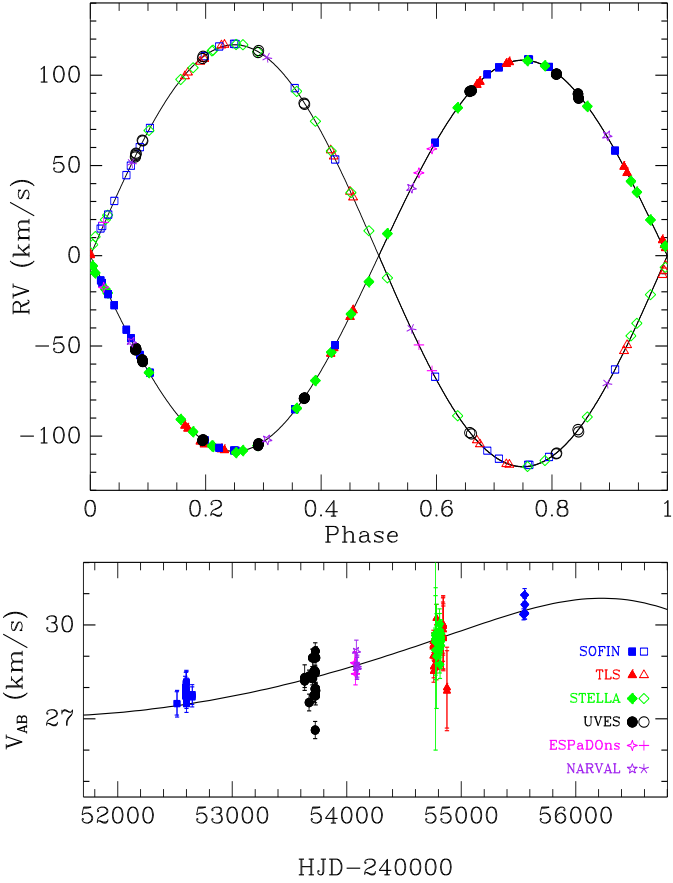}
\caption{
Spectroscopic orbit of the eclipsing system AR\,Aur using SOFIN spectra obtained in 2002 and 2010 (blue squares),
UVES spectra from 2005 (black dots),  TLS-STELLA spectra from 2008/2009 (red triangles and green diamonds),
three ESPaDOnS spectra (pluses), and four NARVAL spectra (stars). 
Open symbols indicate the measurements of the secondary components, which is the less massive star.
The upper panel shows the velocity of the components of the eclipsing pair with respect to the center-of-mass
of the binary; the lower panel shows the systemic velocity of the pair AB with respect to the center-of-mass
of the triple system.
}
\label{fig:araurorbit}
\end{figure}

\begin{table}
\caption{
Orbital and fundamental parameters for the eclipsing system AR\,Aur.
}
\label{tab:orbitaraur}
\centering
\begin{tabular}{cr@{$\pm$}lc}
\hline
\hline
$P$ [d]      & \multicolumn{2}{c}{4.1346657 (adopted)} \\
$T$ (conj)         & 2452848.7073 & 0.0007 \\
$V_{\rm o}$ [km\,s$^{-1}$]      & 28.65 & 0.09 \\
$K_A$ [km\,s$^{-1}$]      & 108.36 & 0.18 \\
$K_B$ [km\,s$^{-1}$]      & 116.92 & 0.17 \\
%$e$      & \multicolumn{2}{c}{0 (adopted)} \\
$a\,\sin\,i$ [$R_\odot$]     & 18.40&0.02 \\
$M\,\sin^3\,i$ [$M_\odot$]  & 4.898 & 0.016 \\
$M_{\rm A}\,\sin^3\,i$ [$M_\odot$]  & 2.542&  0.009 \\
$M_{\rm B}\,\sin^3\,i$ [$M_\odot$]  &2.356&  0.008 \\
$q$                          & 0.9268 & 0.0020 \\
\hline
$v\,\sin\,i_{\rm A}$ [km\,s$^{-1}$]  & 22.87           &      0.08\\
$v\,\sin\,i_{\rm B}$ [km\,s$^{-1}$]  & 22.35           &      0.07\\
$R_{\rm A}\,\sin\,i$ [$R_\odot$] &       1.868  &     0.019    \\
$R_{\rm B}\,\sin\,i$ [$R_\odot$] &       1.825  &     0.018   \\
\hline
$a$ [$R_\odot$]     & 18.41&0.02 \\
$M_{\rm A}$ [$M_\odot$] & 2.544 &  0.009\\
$M_{\rm B}$ [$M_\odot$] & 2.358 &  0.008 \\
$R_{\rm A}$ [$R_\odot$] & 1.799  &     0.013    \\
$R_{\rm B}$ [$R_\odot$] & 1.834  &     0.019   \\
\hline
\end{tabular}
\tablefoot{System velocity $V_{\rm o}$ corresponds to the center-of-mass
velocity of the triple system.} 
\end{table}

The eclipsing system AR\,Aur (HD\,34364, B9V+B9.5V) with an orbital period of 4.13\,d at an 
age of only 4$\times{}$10$^6$\,yr belongs to the Aur OB1 association.
Since its primary star of HgMn peculiarity is exactly on the ZAMS while the secondary is still contracting 
towards the ZAMS (e.g.\ Nordstr\"om \& Johansen \cite{nord1994}),
it presents the best case to study evolutionary aspects of the chemical peculiarity phenomenon.
A presence of a third body in the system was discovered by Chochol et al.\ (\cite{Chochol1988}).
The existence of the as yet unseen third star with a mass of at least 0.51\,M$_\odot$ was
inferred from a light-time effect in the observed minima with a period of 23.7\,yr 
(Albayrak et al.\ \cite{Albayrak2003}, see also Mikul\'a\v{s}ek et al.\ \cite{Mik2010}).
%25--27\,yr.
We re-determined the spectroscopic orbit of this triple system using all spectroscopic data available 
to us (21 SOFIN, 20 TLS, 26 STELLA, and nine UVES spectra). We also used the radial velocities measured by
Folsom et al.\ (\cite{folsom2010}) for three ESPaDOnS and four NARVAL spectra.
In these calculations, we adopted the binary period and parameters of the light-time orbit published by
Albayrak et al.\ (\cite{Albayrak2003}) in order to account for the long-term variation of the center-of-mass
velocity of the eclipsing pair and the light-time effect corrections to the epochs of observations.
Radial velocity curves indicate that the eccentricity is indistinguishable from zero ($e=0.0009\pm0.0011$),
in agreement with the photometric analysis of Nordstr\"om \& Johansen (\cite{nord1994}). We therefore
adopted a circular orbit in our final calculations. 
The orbital parameters are listed in Table~\ref{tab:orbitaraur}  and 
the radial velocity curve is shown in Fig.~\ref{fig:araurorbit}.
It is clear from the lower panel of Fig.~\ref{fig:araurorbit} that the influence of the third
body cannot be neglected in the orbit fitting.

The middle block of Table~\ref{tab:orbitaraur} corresponds to the results of the 
measurements of projected rotational velocities and the lower block to the combination of our
spectroscopic orbit with the photometric inclination and relative radii of 
Nordstr\"om \& Johansen (\cite{nord1994}).
According to these authors, the primary star, which is more massive and hotter, is slightly smaller than
the secondary. However, the spectral lines of the secondary appear slightly narrower,
which, assuming corotation, would suggest a smaller radius. The difference in $v\sin i$ between
both stars, however, is only marginal and, if real, could be explained by
a non-uniform distribution of the elements Fe, Cr, and Ti with latitude.
%used in our analysis. 
%see comment in tex file}
%%FG If really R_B > R_A and vsini_B < vsini_A, the explanation could be:
%%FG that there is a small departure from corotation or that Fe is not uniform in the
%%FG stellar surface in the peculiar star but concentrated slightly toward the equator.
%% However there are some differences between elements that 
Variability of spectral lines associated with a large 
number of chemical elements was reported for the first time for the primary component of this eclipsing 
binary by Hubrig et al.\ (\cite{Hubrig2006a}).
In the  more recent study of this system by Hubrig et al.\ (\cite{Hubrig2010}),
we presented the results of Doppler imaging for the 
reconstruction of the distribution of Fe and Y over the surface of the primary.
We used a spectroscopic time series obtained in 2005 and from 2008 October to 2009 February. 
In the disentangling process, we adopted the photometric flux ratio obtained by 
Nordstr\"om \& Johansen (\cite{nord1994}), i.e.\ the light contributions of stars A and B
to the continuum are adopted as 0.526 and 0.474, respectively.
Our results showed a remarkable evolution of the elemental spot distribution and the overabundances.
Measurements of the magnetic field with 
the moment technique using several elements revealed the presence of a longitudinal magnetic field 
of the order of a few hundred Gauss in both stellar components as well as a quadratic field of the order 
of 8\,kG on the surface of the primary star.

To study the magnetic field geometry on the surface of the components in this system and its correlation
with the surface inhomogeneous element distribution, a number of nights was allocated at the NOT in December 2010.
However, due to bad weather conditions only five polarimetric observations were acquired 
for AR\,Aur. 
The distribution of the observations was fortunately rather random over the rotation/orbital period, 
allowing us to get an idea about the surface element distribution.
The orbit of the system using all spectra of AR\,Aur available to us is presented in Fig.~\ref{fig:araurorbit}.
The rotation phases were calculated using the ephemeris taken from the study 
of Albayrak et al.\  (\cite{Albayrak2003}).

\begin{figure*}
\centering
\includegraphics[width=0.23\textwidth]{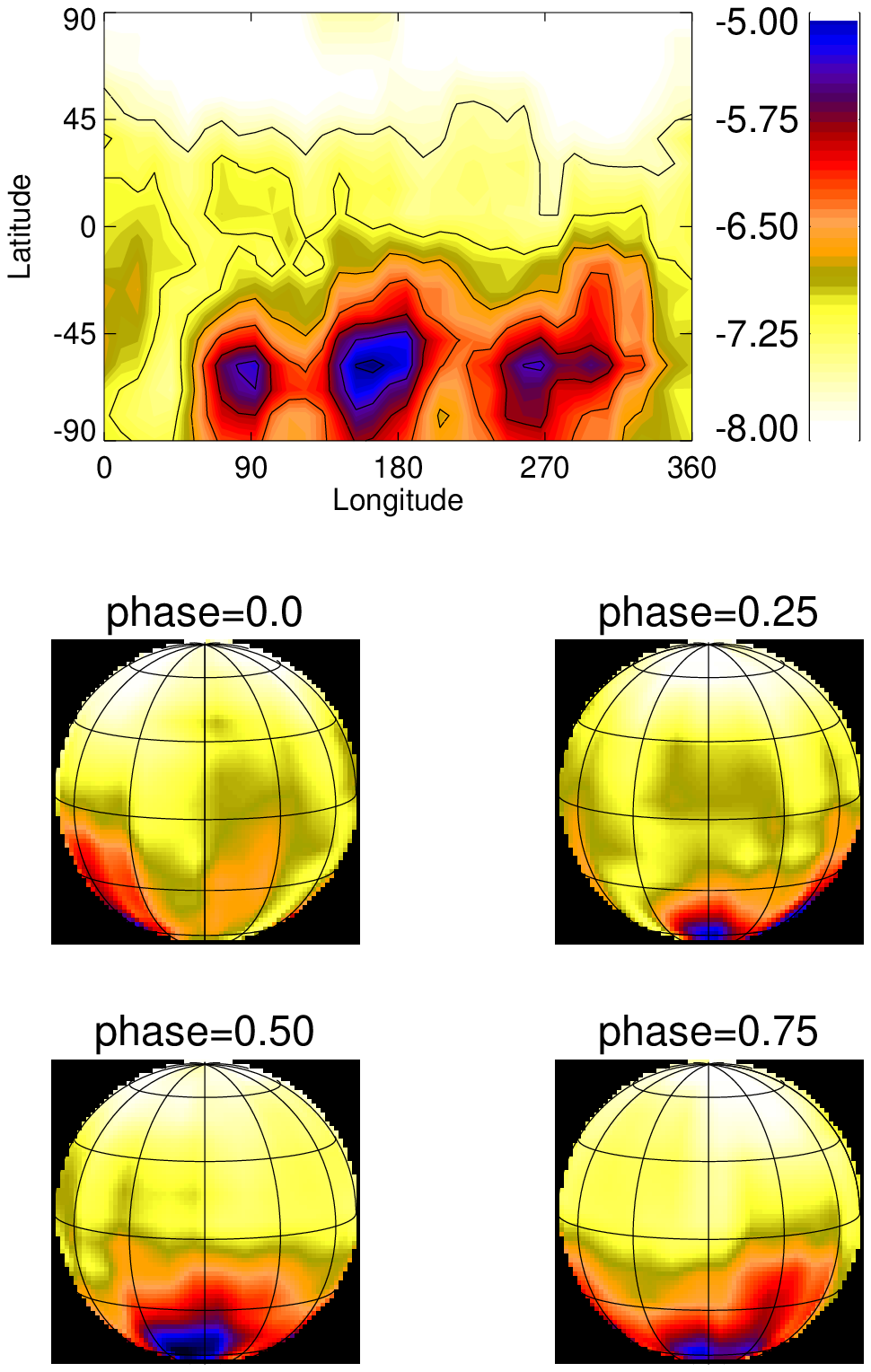}
\includegraphics[width=0.23\textwidth]{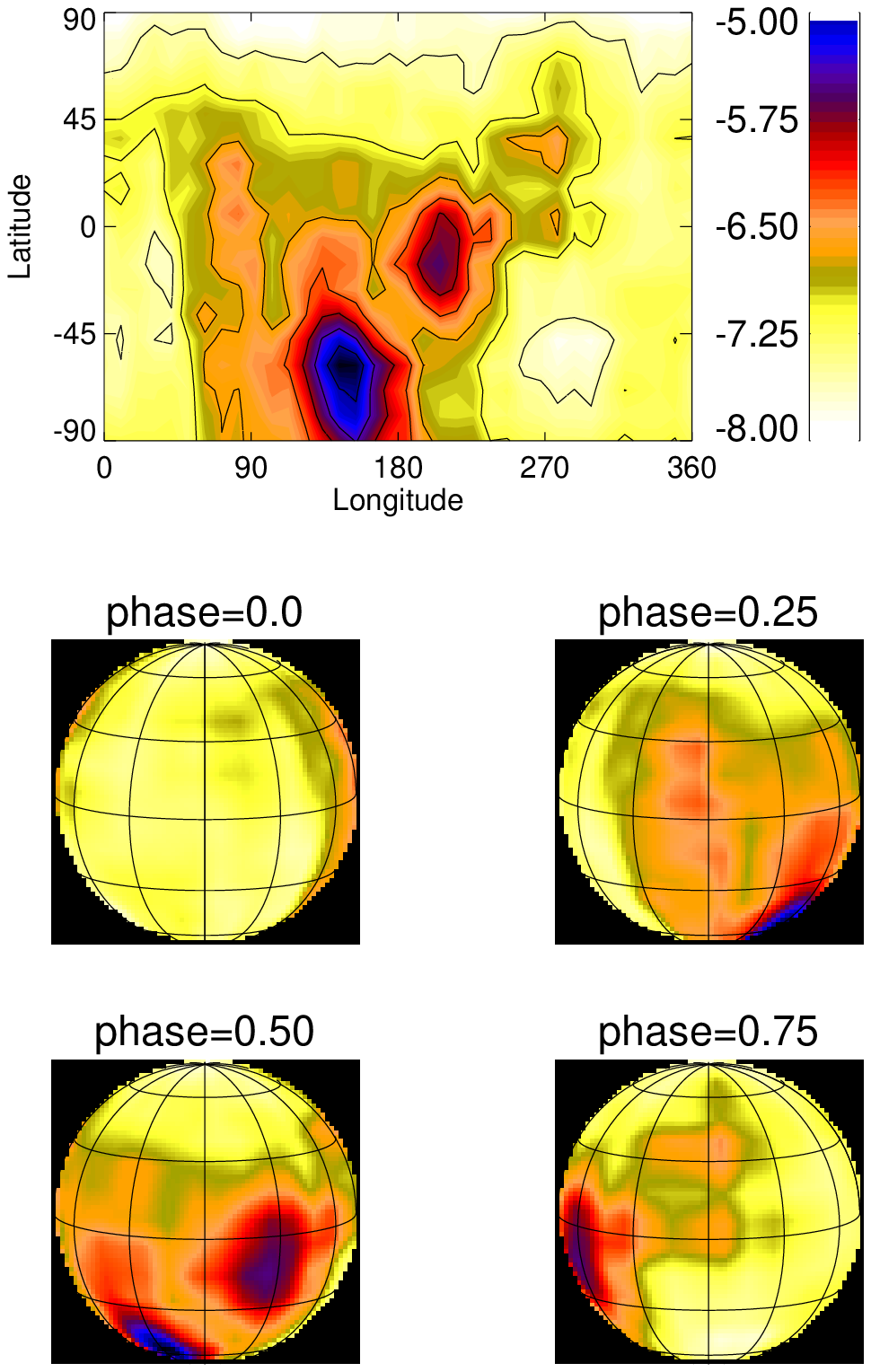}
\includegraphics[width=0.23\textwidth]{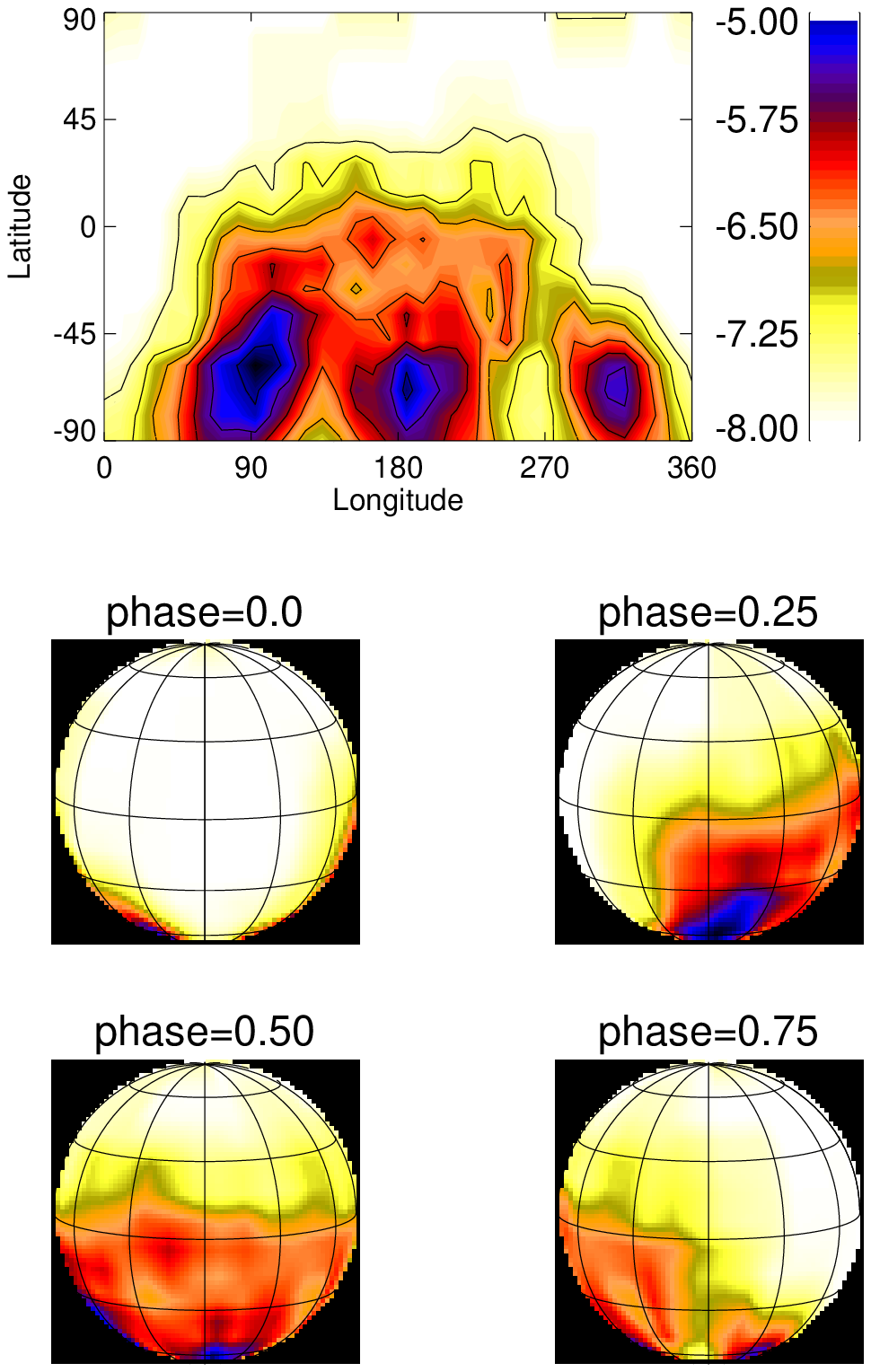}
\caption{
Maps of the abundance distribution for Fe (left), Sr (middle), and Y (right) 
on the surface of the primary in the system AR Aur.
}
\label{fig:doppler}
\end{figure*}

In our analysis, similar to the analysis presented by Hubrig et al.\ (\cite{Hubrig2010}), we used 
the improved Doppler imaging code introduced by Freyhammer et al.\ (\cite{Freyhammer2009}).
This code utilises Tikhonov regularisation with a grid of $6\times6^\circ$ in a way similar to the 
Doppler imaging method described by 
Piskunov (\cite{Piskunov2008}).
In the reconstruction, we searched for the minimum of the regularised 
discrepancy function, which includes the regularisation function and the discrepancy function 
describing the difference between observed and calculated line profiles. All atomic data in our 
analysis were taken from the VALD data base (Kupka et al.\ \cite{Kupka1999}). For the DI reconstruction, we 
selected the following lines belonging to the elements Fe, Sr, and Y: \ion{Fe}{ii} 4923.9\,\AA{},
\ion{Sr}{ii} 4215.5\,\AA{}, and \ion{Y}{ii} 4900.1\,\AA{}.
The computed Doppler maps in Mercator and spherical projections based only on five different rotation 
phases are presented in Fig.~\ref{fig:doppler}. 

\begin{figure}
\centering
\includegraphics[width=0.42\textwidth]{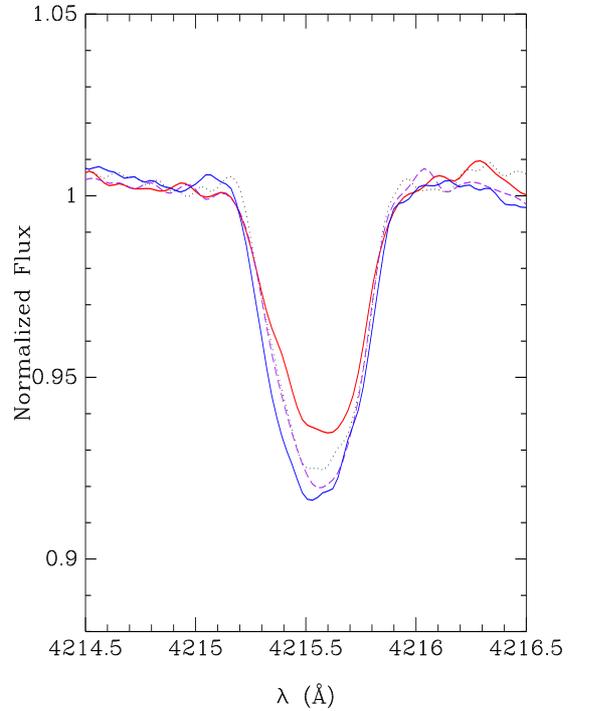}
\caption{
Line profiles of the \ion{Sr}{ii} 4215.5\,\AA{} line around rotation phase 0.68
for all available data sets between 2002 and 2010. 
The thick red continuous line refers to observations with SOFIN in 2010,
the blue continuous line to those with UVES in 2005, the dashed line corresponds to observations
at the Th\"uringer Landessternwarte and STELLA, and the
dotted line corresponds to observations with SOFIN in 2002.
}
\label{fig:sr_profile}
\end{figure}

Clearly, due to the low number of observations these maps cannot be used to study the evolution of 
element distribution over the stellar surface, but they allow us a rough estimate of the 
location of regions with element overabundances. Similar to previous maps obtained for AR\,Aur, we again find that
the regions with lower abundances of  Y and Sr are located on the stellar surface facing the companion,
while spots with Y and Sr high abundances appear on the opposite hemispheres. 
Presently we have at our disposal four spectroscopic data sets obtained between 2002 and December 2010.
The temporal evolution of chemical inhomogeneities is presented in Fig.~\ref{fig:sr_profile}, using as an example
the profile variations of the \ion{Sr}{ii} 4215.5\,\AA{} line.  
All four sets contain spectra at almost the same rotation phases, 0.68. 
The line profiles 
of \ion{Sr}{ii} 4215.5\,\AA{} observed at this phase in all sets show that the Sr spots definitely changed their shape and 
abundance with time (see Fig.~\ref{fig:sr_profile}).
The lowest Sr abundance appeared to be in December 2010, while the 
strongest Sr overabundance was observed in 2005.

\begin{figure}
\centering
\includegraphics[angle=0,totalheight=0.65\textwidth]{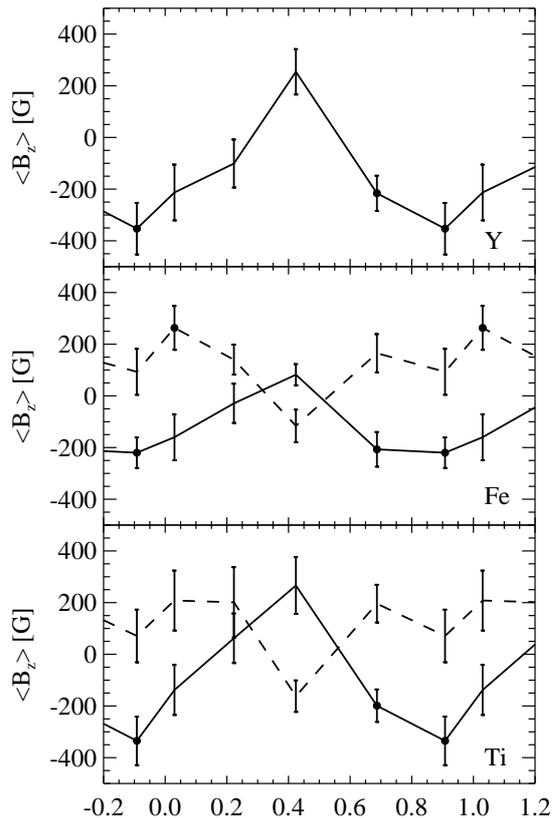}
\caption{
Measurements of the mean longitudinal magnetic field presented as a function of the rotation phase for AR\,Aur.
They were carried out separately for the elements Ti, Fe, and Y (from bottom to top). 
The solid line denotes the primary component, while the dashed line denotes the secondary component.
Filled circles indicate 3$\sigma$ measurements.
}
\label{fig:auromagn}
\end{figure}

In our magnetic field analysis, the longitudinal magnetic field strength measured in AR\,Aur is 
larger than that measured in other targets.
However, due to the much lower resolution of the SOFIN spectra compared with the HARPS spectra and the 
not very high 
S/N, our measurement accuracies are poorest among the values presented in 
Table~\ref{tab:log_meas}. At present, it is 
not clear to us whether the stronger magnetic field in this system can be 
explained by the aspect effect, if we assume that the field topology is characterised by the presence of  
magnetic regions predominantly in the vicinity of stellar equators. Since the AR\,Aur system is an eclipsing binary, 
it is the only system in our sample that offers 
the observer the best visibility of the surface equatorial regions. In any case, this system is one of the most
promising targets to constrain the magnetic field topology in binaries with late B-type primaries.
$3\sigma$ longitudinal magnetic field detections were achieved for measurements using Ti, Fe, and Y 
lines in the spectra of 
the primary in the phases 0.687 and 0.908 as well as for measurements using Fe lines in the spectra of the secondary at the phase 0.030.
The variations of the mean longitudinal magnetic field in both components over the rotation cycle 
presented in Fig.~\ref{fig:auromagn} display a characteristic behaviour similar to that 
found for the binary stars discussed above. The mean quadratic magnetic field is detected at three 
different rotation phases in the primary. 
%Further, crossover was detected at 3$\sigma$ for Fe lines in the phase 0.030.

{\it HD\,53244:}

The SB nature of this target was suggested by Schneider (\cite{Schneider1981}). 
Sch\"oller et al.\ (\cite{Schoeller2010}) found a companion candidate to this star at a separation 
of 0\farcs332 and a position angle of 114.8$^{\circ}$.
The rotation period of HD\,53244, $P=6.16$\,d, was derived by Briquet et al.\ (\cite{Briquet2010}) from the study
of the variations of radial velocities and equivalent widths of spectral lines belonging to 
the inhomogeneously distributed elements Hg and Mn.
However, Y lines appear extremely weak.  No mean longitudinal magnetic field was detected  
using the single HARPS observation of this star. A negative crossover effect at the 3$\sigma$ 
significance level, $-$4073$\pm$1118\,km\,s$^{-1}$\,G, was detected using the sample of Cr lines.

{\it $\kappa$\,Cnc:}

This system is an SB2 with a period of 6.3933\,d,
according to the 9$^{\rm th}$ Catalogue of Spectroscopic Binary Orbits (Pourbaix et al.\ \cite{Pourbaix2009}).
Mason et al.\ (\cite{Mason2001}) found a companion on Besselian year 1999.1606 at a separation 
of 0\farcs286 and a position angle of 108$^{\circ}$.
This star was also resolved by Roberts et al.\  (\cite{Roberts2005}) on Besselian year 2003.0077 with 
a separation of 0\farcs30 and a position angle of 104${^\circ}$.
Sch\"oller et al.\ (\cite{Schoeller2010}) detected this companion at a separation of 0\farcs269 and a 
position angle of 109.7$^{\circ}$.

Two HARPS polarimetric spectra of this system are available in the ESO archive.
The detection of very faint spectral lines of the secondary in both
spectra allowed us to determine  the mass ratio, $q=0.490\pm0.003$. 
Combining our spectroscopic mass ratio with the orbital parameters 
$P$, $e$, and $K_{\rm A}$ of the published SB1 orbit by Aikman (\cite{aikman1976}), we obtain
$M_{\rm A}\,\sin^3\,i = 3.74\pm0.13\,M_\odot$ and 
$M_{\rm B}\,\sin^3\,i = 1.83\pm0.04\,M_\odot$.
To learn about the nature of the secondary component and
measure its rotational velocity, we applied spectral disentangling.
The obtained secondary spectrum
is consistent with a mid-A spectral type, which contributes only
7\% to the total light of the system at $\lambda$5000. 
Since most lines in the spectrum of the secondary component are blended, the rotation
velocity was determined using the technique developed by D\'iaz et al.\ (\cite{rcros}).
%obtaining $v\sin i=42.3\pm1.5$\,km\,s$^{-1}$.
As a result, we obtain $v\sin i=42.3\pm1.5$\,km\,s$^{-1}$.
The primary star presents very narrow spectral lines with the $v\sin i$-value
below 5 km\,s$^{-1}$.
Both components rotate asynchronously. In fact, according to 
estimates of their radii, the $v\sin i$ values for pseudo-synchronisation for stars A and B are
about 33 and 14 km\,s$^{-1}$, respectively, and thus 
very far from the observed values.

With an orbital period of 6.39\,d, the orbital phase difference between the two HARPS observations is 0.41. 
The strongest lines in the spectra belong to the elements Mn and Fe, while Y lines are extremely weak.
The weak negative longitudinal magnetic field is detected at the 3$\sigma$ significance level in the 
HARPS spectrum obtained at MJD\,55202.304 using Fe lines. Also, measurements using Mn lines resulting in 
$\left<B_{\rm z}\right>=-47\pm15$\,G confirm the presence of a weak negative magnetic field at this epoch. 
No significant detection was achieved for the crossover effect and the quadratic magnetic field.
%Ti, Cr not strong. For Mn we measure $-$47$\pm$15 and 15$\pm$15 for both spectra

{\it HD\,101189:}

This target is not known to belong to a SB system.
%There are no references in the literature that indicate multiplicity for this object.
Sch\"oller et al.\ (\cite{Schoeller2010}) found a close companion candidate to HD\,101189 at a separation 
of 0\farcs337 and a position angle of 104.1$^{\circ}$.

The first HARPS spectropolarimetric observation of this star was carried out at MJD\,54982.4744, and its
analysis was reported by Hubrig et al.\ (\cite{Hubrig2011}). 
For the same HARPS observations, Makaganiuk et al.\ (\cite{Makaganiuk2011b}) 
reported a non-detection with  ${\left<B_{\rm z}\right>=-95\pm66}$\,G using the LSD technique.
The quality of the spectra obtained 
at this epoch was especially poor, with achieved S/N between 12 and 52.
The measurements of Hubrig et al.\ delivered the same magnetic field (of the order of $-$200\,G) for both
the regular science Stokes~$V$ observations and the null spectrum. As the authors showed in their Fig.~8, 
this is because the spectrum with the best S/N ratio (52 in that case) significantly 
contributes to the null spectrum. 
This result implies that if the observations are carried out with low S/N, and 
the difference in S/N in the sub-exposures 
is large, the magnetic field cannot be measured conclusively, and there is no use in the null 
spectrum to prove whether or not the detected field is real. 

The second HARPS observation of HD\,101189 was obtained on MJD\,55201.3629.
For this observation, we detect a weak magnetic field at the 3$\sigma$ level
${\left<B_{\rm z}\right>=-74\pm24}$\,G using Ti lines and ${\left<B_{\rm z}\right>=-37\pm11}$\ G using Y lines.
A quadratic field at 3$\sigma$ significance level was detected for all line samples apart from the 
sample of Y lines. The shape of the line profiles of different elements is very different: Y lines appear double
with two absorption maxima in the line profiles, while this profile shape is not observed in Mn, Fe, Cr, 
and Ti lines, suggesting a different spot distribution of these elements over the stellar surface. Only for the sample
of Y lines do we observe the presence of a weak positive crossover effect, 764$\pm$338\,km\,s$^{-1}$\,G, 
at the 2.3$\sigma$ level.

{\it HD\,221507:}
This target is not known to belong to a SB system.
Hubrig et al.\ (\cite{Hubrig2001}) studied this target with ADONIS and did not
detect any companion.
However, Sch\"oller et al.\ (\cite{Schoeller2010}) found a companion candidate to HD\,221507 at a 
separation of 0\farcs641 and a position angle of 240.2$^{\circ}$.

The rotation period of this star, $P=1.93$\,d was derived by Briquet et al.\ (\cite{Briquet2010}) from the study
of the variations of radial velocities and equivalent widths of spectral lines belonging to 
the inhomogeneously distributed elements Hg, Mn, and Y.
We detect a weak longitudinal magnetic field at 3$\sigma$ level ${\left<B_{\rm z}\right>=78\pm25}$\,G using Y lines.
Also, measurements using Mn lines, ${\left<B_{\rm z}\right>=75\pm23}$\,G,  show the presence of a positive 
longitudinal magnetic field. No significant crossover effect and quadratic magnetic field are detected in our
measurements.
%Ti, Cr, Mn, Fe, Y are strong, For Mn we measure 75$\pm$23\,G.

{\it HD\,179761 and HD\,209459:}

The star HD\,179761 is considered in the literature as a normal B-type star and HD\,208459 
as a superficially normal B-type star. Superficially normal B-type stars 
are generally 
indistinguishable from the normal stars in terms of their iron-peak elemental abundances (e.g.\ 
Smith \& Dworetsky \cite{Smith1993}). They are usually sharp-lined stars that possess normal MK 
classifications, but nonetheless exhibit mild peculiarities when observed at high spectral resolution.

Although HD\,209459 is not a typical HgMn star, it is listed in the catalogue of HgMn stars 
by Schneider (\cite{Schneider1981}).
Our analysis of the HARPS spectropolarimetric material for this star indicates the absence of Hg and Y 
lines and the presence of weak Mn lines. The lines of Ti, Cr, and Fe appear rather strong and reveal a 
weak variability.

 Ti, Cr, Mn, and Y lines are weak in the HARPS spectrum of HD\,179761. We
cannot report anything on the variability of this star as only a single HARPS observation is available.

Hubrig et al.\ (\cite{Hubrig1999b}) and Hubrig \& Castelli (\cite{HubrigCastelli2001}) used an anomalous 
strength of the \ion{Fe}{ii} $\lambda$6147.7 
line relative to the \ion{Fe}{ii} $\lambda$6149.2 line in stars with magnetic fields to study the presence 
of magnetic fields in HgMn, normal, and superficially normal late-B type stars. In this method, previously 
introduced by Mathys (\cite{Mathys1990}), the observed relative 
differences between the equivalent widths of the two 
\ion{Fe}{ii} lines are compared with those derived from synthetic spectra computed by neglecting
magnetic field effects. While the differences between the equivalent widths for HD\,209459
were found to lie within the error limits, the differences for HD\,179761 were found to be much larger,
even if all observational and computational uncertainties are taken into account 
(Hubrig \& Castelli\ \cite{HubrigCastelli2001}), which  suggests the presence of a magnetic field in this star. 
Both HD\,179761 and HD\,209459 were also observed with FORS\,1 at the VLT by Hubrig et al.\ (\cite{Hubrig2006b}),
who reported a 3$\sigma$ detection for the mean longitudinal magnetic field of HD\,179761.

The observations with HARPS indicate the presence of a weak longitudinal magnetic field
in HD\,209459 at MJD=55417.149, measured using the samples of Ti, Cr, and Fe lines.
A quadratic magnetic field of the order of 3\,kG at the 3$\sigma$ level was detected in HD\,179761 using Fe lines,
and much weaker quadratic fields of the order of 1.3--1.5\,kG were detected in both observations of 
HD\,209459 using Cr lines. The presence of a $\sim$3\,kG quadratic field in
HD\,179761 agrees well with the previous analysis of the relative magnetic intensification of the two
\ion{Fe}{ii} lines by Hubrig \& Castelli (\cite{HubrigCastelli2001}).
We note that it is not the first time that a quadratic magnetic field is detected in upper main sequence
stars considered as normal or superficially normal stars. Mathys \& Hubrig (\cite{Mathys2006}) studied the 
presence of a quadratic magnetic field in the superficially normal A-type star HD\,91375. Their analysis 
led to the detection of a 1.7\,kG quadratic field at a significance level of more than 4$\sigma$.  

%In HD\,179761 all Mn and Y, Cr and Ti are weak, quadratic field of $\sim$3\,kG is in 
%accordance with the previous study of Hubrig \& Castelli (\cite{HubrigCastelli2001}).
%HD\,209459: No Y, Mn is weak, Fe, Ti and Cr are strong.
%In our spectra the \ion{Ti}{ii}, \ion{Cr}{ii}, and \ion{Fe}{ii} lines appear slightly variable.
% lines are definitely variable in HD\,209459 (4183.648, 4386.8, 4563.8, 4805.1, 4911),
%\ion{Cr}{ii} (4616, 4618), \ion{Fe}{ii} lines (4273, 4923, 5506, 6456).

\section{Discussion}
\label{sect:disc}

The region of the main sequence centred on A and B stars, 
also referred to as the ``tepid stars'', represents an ideal laboratory to 
study a wide variety of physical processes that are at work to a greater or 
lesser extent in most stellar types. These processes include radiation driven 
diffusion, differential gravitational settling, grain accretion, magnetic 
fields and non-radial pulsations.  
Understanding them is becoming increasingly important for the refinement of stellar 
evolution models and for the improved treatment of the stellar contribution in
studies of galactic evolution. 
Presently, the radiative diffusion hypothesis developed by Michaud (\cite{Michaud1970}) is the most 
frequently accepted theory for producing the observed anomalies in HgMn stars.
According to Michaud et al.\ (\cite{Michaud1974}), the effect of diffusion would cause a cloud of mercury 
to form high up in the HgMn star atmospheres. However, the impact of the radiative diffusion process 
has not yet been studied for any element considered in our work. 
%Therefore, the mechanisms responsible for 
%producing the chemical peculiarity of numerous elements are still unclear.

Based mostly on ESO/HARPS archive data, our analysis suggests the existence of 
intriguing correlations between magnetic field, abundance anomalies, and binary 
properties. In the SB2 systems with the synchronously 
rotating components, 41\,Eri and AR\,Aur, the stellar surfaces facing the companion star usually 
display low-abundance element spots and negative magnetic field polarity. 
The surface of the opposite hemisphere, as a rule, is covered by high-abundance
element spots and the magnetic field is positive at the rotation phases of 
the best-spot visibility. 
Also, Measurement results for the SB1 system HD\,11753 indicate that element
underabundance (respectively overabundance) is observed where the polarity of
the magnetic field is negative (respectively positive). 
In the case of the very young pseudosynchronously rotating system 66\,Eri, additional future observations 
are urgently needed
to determine the period of the spectral variability and to investigate the presence of element-spot 
evolution.
It is quite possible that because the system is not yet synchronised, a dynamical spot evolution
is more pronounced compared to synchronised systems.

Although only rather weak longitudinal magnetic fields
are detected on the surface of our targets, the numerous 3$\sigma$ detections of mean quadratic magnetic fields
strongly suggest that magnetic fields are present in their atmospheres. We note again the advantage
of carrying out quadratic magnetic field measurements as they
provide measurements of the magnetic field strength that are somewhat insensitive to the magnetic field structure,
and especially suited for magnetic fields with complex structure.
%Thus it is especially well suited to the  
%detection of fields that have a complex structure, as is likely the case in our targets. 

The interactions of differential rotation, magnetic fields, and 
meridional flows in a tidal potential may lead to processes that
can explain complex surface patterns. Possible instabilities
should be studied by non-linear, three-dimensional simulations.
The resulting fluctuations in velocity and magnetic field may
not only explain the relatively small scales of surface magnetic fields,
but also provide enhanced transport of angular momentum that explain
the high rates of synchronisation.
Synchronisation in an orbit with about 5\,d period, an initial
rotation period of 1\,d , and stellar parameters similar to 66~Eri is
roughly $10^4$ times the dissipation time scale in the star (Zahn \cite{Zahn2008}).
In a non-turbulent, non-magnetic star where only the microscopic gas 
viscosity can couple the surface with the interior, it would be 
impossible to reach synchronisation within the age of the universe.
There are also gravity waves that lead to synchronisation, but for 
stars of 2--3~solar masses, this effect still takes roughly $10^9$~years.
Large-scale magnetic fields or turbulence can provide enough coupling
to reach (pseudo-)synchronisation in about a million years.

\begin{figure}
\centering
\includegraphics[angle=270,width=0.45\textwidth]{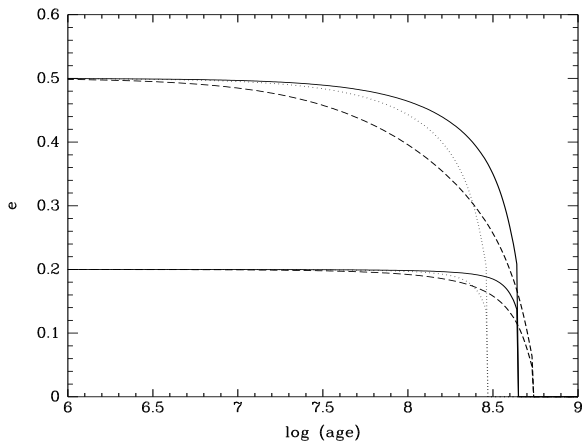}
\caption{
Circularisation times for 41\,Eri (dotted lines), 66\,Eri (solid lines),  and AR\,Aur (dashed lines), 
assuming that they formed with eccentric orbits with $e_0$ = 0.2 and 0.5.
}
\label{fig:et}
\end{figure}

As an illustration, we calculated the circularisation times for the systems 41\,Eri, 66\,Eri, and AR\,Aur, 
assuming that they formed with eccentric orbits.
With the present parameters of the stars, the circularisation time scales, 
$ (-d\ln e/dt)^{-1}$ could be as high as $10^{10}$ yr. As they evolve within the 
main sequence, circularisation becomes more efficient, but even so they will 
arrive at the terminal age main sequence (TAMS) with a significant fraction of their original eccentricity: 
$e/e_0$=0.15--0.70, depending on the original eccentricity $e_0$.
Figure~\ref{fig:et} shows the evolution of $e$ for the three systems, starting from $e_0$ = 0.2 and 0.5.
The solid line corresponds to 66\,Eri, the dotted line to 41\,Eri, and the dashed line to AR\,Aur. 
The  vertical end of the curves to the right mark the TAMS.
The calculations were performed using the stellar models calculated by Claret (\cite{Claret2004}), 
which include the internal structure constants required to calculate
the orbital evolution.
%I used eq. 1 of Claret (2011 A&A 526, A157) taken R(t), L(t), k2(t) from the 
%Claret (2004) stellar models
Considering that all three systems are young, with ages well below $10^8$ yr, 
it would be expected that their present eccentricities woould not differ 
significantly from the original ones. The observed very low or null eccentricities 
cannot be explained as a consequence of standard tidal evolution. 

The magneto-rotational instability is a candidate that can provide
turbulence for a wide range of magnetic field strengths and field
topologies, but only if the internal angular velocity of the star 
decreases with the distance from the rotation axis.
%decreases (Omega nimmt mit dem Achsabstand ab) from the inside to the
%outside of the star. 
Since synchronisation
in the systems considered here is in fact a braking process, the
star may exhibit a slower rotation near its surface than deeply
inside. Numerical simulations should elucidate the expected enhancement
of the viscosity in the star and will also deliver surface topologies
of the fields resulting from this or other instabilities that can be
compared to the observed patterns.
%Clearly, numerical simulations of magnetic stellar configurations have to 
%complement the presented observations. The interactions of differential rotation,
%magnetic fields and meridional flows in a tidal potential should be studied
%in nonlinear, three-dimensional simulations. They should in particular focus on the 
%coupling of the outer layers of the stars by magnetic fields and/or
%turbulence, which is supposed to explain both the observed high rates 
%of synchronisation and the complex surface pattern

Tidal forces strongly depend on the distance between the components.
For the three SB2 systems with synchronised and pseudosynchronised  components, 41\,Eri, 66\,Eri, and AR\,Aur, we estimated the 
fractional radii $r_j=R_j/a$.
% in the following way:
Assuming for 41\,Eri

\begin{eqnarray}
(v\,\sin\,i)_A	&=& 12.23\pm{}0.06\,{\rm km/s}, \nonumber \\
(v\,\sin\,i)_B &=& 11.78\pm{}0.07\,{\rm km/s}, \nonumber \\
% (v\,\sin\,i)_A	&=& (v\,\sin\,i)_B = 12.2\pm{}0.1\,{\rm km/s}, \nonumber \\
 K_A+K_B &=& 127.38 \pm 0.23\,{\rm km/s}, \nonumber \\
 e&=&0,\nonumber
\end{eqnarray}
it follows that $R_A/r = 0.096\pm{}0.001$ and $R_B/r = 0.093\pm 0.001$

If we assume for 66\,Eri pseudosynchronisation
($P_{\rm rot} = 5.3015\,d$), we obtain $R_A/r = 0.084\pm 0.001$
and  $R_B/r = 0.080\pm 0.001$.

For AR\,Aur we use
\begin{eqnarray}
 (v\,\sin\,i)_A &=& 22.87\pm 0.08\,{\rm km/s}, \nonumber \\
(v\,\sin\,i)_B &=& 22.35\pm{}0.07\,{\rm km/s}, \nonumber \\
% (v\,\sin\,i)_B &=& 22.25 \pm 0.30\,{\rm km/s}, \nonumber \\
 K_A+K_B &=& 225.28 \pm 0.18\,{\rm km/s}, \nonumber \\
 e &=& 0 \nonumber
\end{eqnarray}
and obtain  $R_A/a = 0.102\pm 0.001$ and $R_B/a = 0.099\pm 0.001$.\\

Since the fractional radii in these systems are quite large, 
the components in the close binary systems are subject to tidal forces
acting differentially throughout their bodies and causing various
effects in the system. The tidal effects include synchronisation of the rotational periods 
with the orbital period, circularisation of the orbit, and  
alignment of the rotation axes with the orbital axis. All these 
changes in the dynamical properties of the stars happen on different
time scales. 
The spin-down may be a surface effect, but the star then possesses
an internal differential rotation that may lead to hydrodynamic
or magnetohydrodynamic instabilities. Even a moderately turbulent
state generated by an instability will lead to much stronger coupling
with the stellar interior and redistribute angular momentum.
%Clearly, any physical mechanism considered 
%responsible for the generation of element spots and the underlying weak magnetic fields, has to 
%include the presence of rather strong tidal interactions.

\begin{figure*}
\centering
\includegraphics[angle=270,totalheight=0.55\textwidth]{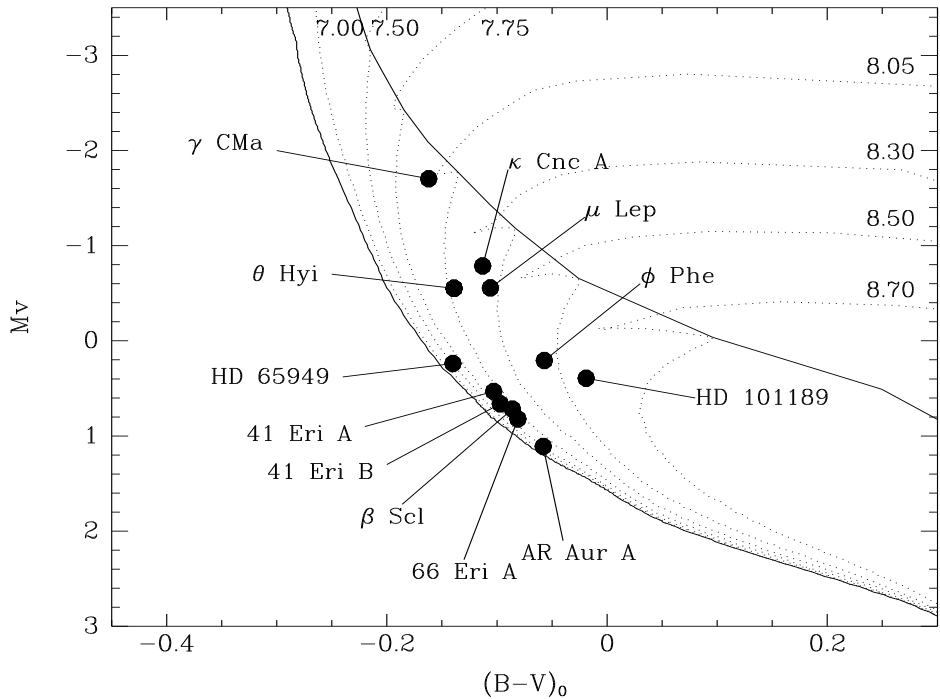}
\caption{H-R diagram for the HgMn stars in our sample. Continuous lines present the ZAMS and TAMS. Dotted lines 
present isochrones for log(t) from 7.00 to 8.70.
%7.00, 7.50, 7.75, 8.05, 8.20, 8.50, and 8.70.
}
\label{fig:HR}
\end{figure*}

To establish the evolutionary status of the HgMn stars in our sample, we studied their 
distribution in the H-R diagram.
To this aim, the distances were calculated mainly using Hipparcos parallaxes
(van Leeuwen \cite{hipp}). 
For stars with high galactic latitudes, 
%(HD19400, HD27376, HD32964, HD33904, HD221507, ..)
we estimated the interstellar absorption using 
Schlegel et al.\ (\cite{Schlegel1998}) maps and the distance by applying the
same procedure as Bilir et al.\ (\cite{Bilir2008}). 
%this involve: LK correction of the parallaxes (eq. 1 of Bilir paper),
%the calculation of the total galactic reddening in the object direction
%using the Schelegel maps (http://ned.ipac.caltech.edu/forms/calculator.html), 
% the calculation of the actual interstellar absorption using
% the object distance and galactic latitude using eq. 3 in Bilir.
Since this approach is not valid for targets located very close to the galactic plane,
we estimated the absorption in a different way for the low-latitude stars HD\,101189 and AR\,Aur..
HD\,101189 is located in the field of the open cluster NGC\,3766, but it is a foreground star.
This fact allowed us to estimate the absorption to the star using 
cluster reddening and the distance ratio.
% NGC3766 has E(B-V)=0.175 with d=1750 pc
% HD101189 has d=88pc, so E(B-V)=0.175/1750*88=0.009 mag, Av=0.027  
For the eclipsing binary AR\,Aur, we adopted the absolute stellar parameters derived
by Nordstrom \& Johansen (\cite{nord1994}) from light and radial velocity curves. 
%The peculiar star is the most massive, hotter, and slightly smaller companion. 
%These authors give Mv_1 = 1.11 +- 0.13 mag, and reddening=0   
%An alternative is to use the Hipparcos parallax and E(B-V)=0, following Bilir et al (2008).
%Bilir et al.\ (2008) estimated, using the Hipparcos parallax, an absolute magnitude
%0.25 mags fainter, which would place this star slightly below the ZAMS.
The star  HD\,65949, which is the faintest and most distant star, belongs to
the open cluster NGC\,2516. We used, therefore, the cluster distance and
reddening published by Terndrup et al.\ (\cite{phot2516}) to obtain absolute magnitude and intrinsic color.
% I used V=8.38 B-V=-0.02 and the cluster parameters of Terndrup et al (2002):
% (m-M)o=7.93+-0.14 and E(B-V)=0.12.
% An alternative calculation is the use of Hipparcos parallaxes of other stars
% in the cluster. This gives a slightly shorter distance. It would appear 0.17 mags 
%fainter and 0.02 mag redder in the diagram, being also on the ZAMS.
%I note that the B-V index in the cluster photometric studies is 
% between -0.04 and -0.01 (average -0.02), while in simbad it is -0.06. 
%this difference is will noticeable in the HR diagram. It puts it out of the MS.
In the case of double-lined spectroscopic binaries, the position of the individual
components in the H-R diagram was calculated using the spectroscopic
mass ratio, from which the magnitude difference between the components
was estimated by interpolating in the Geneva stellar models (Schaller et al.\ \cite{Schaller1992}).
%HD78316: Mv_2-Mv_1 = 2.83 mag Primary is the HgMn star
%HD27376: Mv=-0.157mag, (B-V)o=-0.100; both companions are HgMn
%         q=0.969 => Mv_2-Mv_1 = 0.13 mag; (B-V)_2-(B-V)_1=0.006
%                => Mv_1=0.532   Mv_2=0.662  (B-V)__1=-0.103  (B-V)_2=-0.097
%HD32964: Mv=0.111 (B-V)o=-0.078 q=0.975 (the HgMn star is the less massive)
%	  Mv_2=0.908  (B-V)_2=-0.071
Figure\,\ref{fig:HR} shows the position of the 11 HgMn stars in the color-magnitude
diagram, together with isochrones taken from Schaller et al.\ (\cite{Schaller1992}).  
Six stars in our sample, 41\,Eri\,A, 41\,Eri\,B, 66\,Eri\,A, AR\,Aur\,A, HD\,65949, and HD\,221507, are located very close 
to the ZAMS. 
%HD\,65949, HD\,27376A, HD\,27376B, HD\,221507, HD\,32964B, and AR\,Aur.
%HD\,101189 seems to be significantly older than the others.
 
From the knowledge of orbital parameters, fundamental parameters of components, and
fractional radii, 
it is possible to estimate the reflection effect, gravitational distortion, and darkening
for a typical HgMn star in a SB2 system.
One of the consequences of the proximity of the stars is the departure from sphericity 
of the stellar surface. The star acquires a somewhat larger dimension along the line passing 
through the two star centres, while it has a smaller radius in the direction that is perpendicular to the 
former line and also lies in the orbital plane.
The relative differences between both radii are 0.09\% and 0.08\% for the two components 
of the system 41\,Eri, 0.05\% and 0.06\% for the system 66\,Eri, and 0.09\% and 0.10\% for the system AR\,Aur. 
Observational effects of such stellar shapes are, for example, the variation of the 
$v\sin i$-value, which in the three systems studied is below the observational 
errors, and light variations due to the variation of the stellar disk size, which 
is expected to be of the order of 2--3\,mmag.
An effect that could affect physical conditions in the stellar atmosphere is the 
reflection effect. Assuming an albedo equal to 1 (the usual assumption for stars with 
radiative envelopes) and the temperatures and geometrical configuration of each system, 
we estimate that the temperature excess in the surface region facing the 
companion is between 20 and 40 K, which represents 0.18--0.38\% of the star temperature.
The question to solve is whether such small temperature differences can be considered a reasonable explanation 
of the origin of the low- and high-abundance spots observed either on the surface facing the companion 
or on the surface hidden to the 
companion. Although it is currently not easy to give an answer without a specific model of 
time-dependent diffusion processes, it seems very unlikely 
that these processes are the main cause of the observed abundance pattern.
We note that the early circularisation and a very specific configuration of chemical spots in binary systems
support the idea of an important impact of magnetic fields on the physical processes taking place in these stars.
Since the orbital parameters of close binaries 
can change as a consequence of the action of tidal forces or other
mechanisms, the influence of the presence of a stellar companion on the
physical processes playing a role in the formation of chemical anomalies
is expected to vary with time. 
The atmospheric physical conditions for the
development of chemical inhomogeneities might appear favourable at certain stages of the
stellar life and unfavourable at other stages.
%according to the binary companion) would support the presence of magnetic fields in 
%these stars.
%Even though it is not easy to give an answer without a specific model of 
%the dependence of the diffusion processes on the temperature, it seems very unlikely 
%that this is the main cause of the observed abundance spots.
%In conclusion, both facts (early circularisation and configuration of chemical spots 
%according to the binary companion) would support the presence of magnetic fields in 
%these stars.
% Our estimates show that for 
%binaries with  periods of about three days and large mass ratio $q$ (close to 1), the reflection effect is 
%smaller that 1\% in the flux and temperature, i.e.\ $\Delta$T$\sim$100\,K. For low and intermediate $q$-values
%the gravitational darkening and distortion become more important than reflection. 

%The inclination angle $i$ sample different values. Since the information of the predominant distribution
%of different elements over the surface in HgMn stars is rather poor, it is important to sample
%different inclination angles.
Further studies using high S/N circularly polarised spectra well-distributed throughout 
the rotation period of a representative sample of  HgMn spectroscopic binaries 
sampling a wide range  of inclinations of the rotation axis to the line of sight are
urgently needed to map their 
magnetic field and elemental abundances with a Zeeman-Doppler imaging code.
The sampling of a wide range  of inclination of the rotation axis to the line of sight
in HgMn systems is especially important because the latitudinal information on the surface abundance distribution
of different elements in HgMn stars is still rather poor.
In this way, it will be possible to probe  
correlations between the binary properties, the magnetic field structure, and 
the abundance inhomogeneities.
At the same time, we will be able to gain insight into the mechanisms
responsible for the chemical inhomogeneities and their dynamical evolution on the surface of HgMn stars.
%as well as shed light on the origin of the inconsistencies between the results of 
%magnetic field detection attempts of different groups. 
%Such studies will 
%contribute to the understanding of the various physical mechanisms that are at 
%play in the outer layers of stars of most types, whose most extreme
%manifestations are observed in A- and B-type main sequence stars.

{
\acknowledgements
We are grateful to G.~Mathys for valuable discussions and to the anonymous referee for useful comments. 
This research made use of the SIMBAD database,
operated at CDS, Strasbourg, France. 
HK acknowledges the support from the European Commission 
under the Marie Curie IEF Programme in FP7. 
Nordic Optical Telescope is operated on the island of La Palma jointly by  Denmark, Finland, Iceland, 
Norway, and Sweden, in the Spanish Observatorio del Roque de los Muchachos of the Instituto de 
Astrofisica de Canarias.
}

%{\bf there are a few corrections and references added}

\end{document}